\documentclass[aps,pra,twocolumn,superscriptaddress]{revtex4-2} 
\usepackage{graphicx}  
\usepackage{epstopdf}
\usepackage{dcolumn}   
\usepackage{bm}        
\usepackage{amsmath}   
\usepackage{amssymb}   
\usepackage{mathtools}
\usepackage{maybemath}
\usepackage{xcolor}
\usepackage{enumitem}
\usepackage{graphicx}  
\usepackage{nicefrac}
\usepackage{comment}
\usepackage{physics}
\newcommand{\eps}[0]{\varepsilon}

\newcommand{\up}[0]{\uparrow}

\newcommand{\dw}[0]{\downarrow}
\newcommand{\ho}[0]{\mathrm{ho}}

\newcommand{\lp}[0]{\left}
\newcommand{\rp}[0]{\right}
\newcommand{\s}[0]{\sigma}
\newcommand{\sbar}[0]{{\bar{\sigma}}}
\newcommand{\de}[0]{\partial}
\newcommand{\KS}[0]{\mathrm{KS}}

\newcommand{\leqs}[0]{\leqslant}

\newcommand{\half}[0]{\frac{1}{2}}
\newcommand{\thalf}[0]{\tfrac{1}{2}}
\newcommand{\rr}[0]{\mathbf{r}}

\newcommand{\m}[0]{\textrm{min}}

\usepackage{xr}
\newcommand\identity{1\kern-0.25em\text{l}}

\begin{document}

\title{Spin migration in density functional theory: energy, potential and density perspectives}

\author{Alon Hayman}
\affiliation{Fritz Haber Research Center for Molecular Dynamics and Institute of Chemistry, The Hebrew University of Jerusalem, 9091401 Jerusalem, Israel}

\author{Eli Kraisler}
\email[Author to whom correspondence should be addressed: ]{eli.kraisler@mail.huji.ac.il}
\affiliation{Fritz Haber Research Center for Molecular Dynamics and Institute of Chemistry, The Hebrew University of Jerusalem, 9091401 Jerusalem, Israel}

\author{Tamar Stein}
\email[Author to whom correspondence should be addressed: ]{tamar.stein@mail.huji.ac.il}
\affiliation{Fritz Haber Research Center for Molecular Dynamics and Institute of Chemistry, The Hebrew University of Jerusalem, 9091401 Jerusalem, Israel}
\date{\today}

\begin{abstract} 

Spin is a fundamental property of any many-electron system. The ability of density functional theory to accurately predict the physical properties of a system, while varying its spin, is crucial for describing magnetic materials and high-spin molecules, spin flip, magnetization and demagnetization processes. Within density functional theory, when using various exchange-correlation approximations, the exact dependence of the energy on the spin often deviates from the exact constant or piecewise-linear behavior, which is directly related to the problem of strong (static) correlation and challenges the description of molecular dissociation. In this paper, we study the behavior of the energy, the frontier Kohn-Sham (KS) orbitals, the KS potentials and the electron density, with respect to fractional spin, in different atomic systems.
We analyze five standard exchange-correlation functionals and find three main scenarios of deviation from the expected exact results. We clearly recognize a jump in the frontier orbital energies upon spin variation in the exact exchange and in hybrid functionals, and the related plateau in the corresponding KS potential.  Our results are instrumental for the assessment of the quality of existing approximations from a new perspective and for the development of advanced functionals with sensitivity to magnetic properties. 

\end{abstract}

\maketitle{}

\section{Introduction}

Density functional theory (DFT)~\cite{KS65,PY} is a powerful method for describing the electronic structure in materials 
and is widely used in chemistry, physics and materials science~\cite{Maurer19,tsuneda2014density,Giustino14_materials,ShollSteckel11,huang2008advances,Cramer04}. 
Within the Kohn-Sham (KS) approach to DFT, 
an interacting many-electron system is described via an auxiliary system of non-interacting electrons, which are subject to an effective potential, chosen to reconstruct the density of the interacting system. 
In principle, DFT is an exact approach, but in practice, the aforementioned effective potential, particularly its exchange-correlation (XC) component, is unknown and needs to be approximated.

During the last six decades, many XC approximations have been suggested, with varying degrees of accuracy for different systems and physical properties. In parallel, several approaches for developing XC functionals have emerged. In particular, the approach of satisfying exact constraints~\cite{PerdewSchmidt01,Perdew09_perplexed,KaplanLevyPerdew23,Sun_SCAN_PRL15,Saavedra12} proved especially useful if one aims at a global, first-principles functional that is relevant for a variety of materials: Although the exact XC functional is unknown, many of its properties can be derived analytically or observed numerically for small systems, relying on highly-accurate calculations.

Two exact conditions relevant to this study are the well-known piecewise-linearity condition with respect to fractional electron number, and the constancy condition with respect to fractional spins. The piecewise-linearity condition states that the exact total energy versus particle number is a series of linear segments between integer electron points with a possible change in the slope when crossing an integer number.\cite{PPLB82} The constancy condition with respect to fractional spin states that the total energy is constant for systems with constant charge but varying spins.\cite{CohenMoriSYang08,Cohen08} The combination of the two exact constraints leads to the flat plane condition.\cite{MoriS09} Notably, these conditions are completely general quantum principles and apply to any many-electron approach.

The occurrence of fractional spins and their effect has been extensively studied in the literature.\cite{CohenMoriSYang08,Cohen08,cohen12,su2018describing,MoriS09,cohen_fractional_MP2_09,burgess2023dft+,BajajKulik17,BajajKulik22,de2021quantifying,Prokopiu22,WodynskiArbuzKaupp21,XDYang16,Yang00,Chan99} In Ref.~\cite{CohenMoriSYang08}, the authors showed that the exact total energy of a system with a fractional $S$ must be constant for all $S \in [-S_\m, S_\m]$ (where $S_\m$ is the \emph{equilibrium} value of the \emph{total} spin, i.e., that value for which the system has the lowest energy), Importantly, it has been demonstrated that errors in fractional spins are related to large static correlation errors in chemical systems, when calculating chemical bond dissociation processes and band structure of Mott insulators.
In Ref.~\citenum{MoriS09} the \emph{flat-plane condition} has been developed, unifying piecewise-linearity~\cite{PPLB82} and the constancy condition~\cite{CohenMoriSYang08}. Focusing on the H atom, Ref.~\citenum{MoriS09} described its exact total energy, while continuously varying $N$ from 0 to 2 and $S$ accordingly, while $S \in [-\tfrac{1}{2}, \tfrac{1}{2}]$. The exact energy graph of H in the $N_\up - N_\dw$ plane is comprised of two triangles whose vertices lie at the points of integer $N_\up$ and $N_\dw$. 

Recently, in Ref.~\cite{GoshenKraisler24} the exact ground state of a general, finite, many-electron system with an arbitrary fractional
electron number, $N$, and fractional spin, $S$, has been rigorously described, using the ensemble approach, and generalizing the piecewise-linearity principle and the flat-plane condition. It has been found which
pure ground states contribute to the ensemble, and which do not. In particular, it has been shown that the contributing pure states are not necessarily among the four neighboring states to the point $(N,S)$ on the $N_\up - N_\dw$ plane, contrary to previous findings; the energy graph consists not only of triangles but also of trapezoids. 

In parallel, Ref.~\cite{BurgessLinscottORegan24} described the energy profile versus electron number and spin for finite systems, also beyond the range $[-S_{\min},S_{\min}]$, relying on the infinite-separation technique~\cite{Yang00}. It has been shown that the energy profile satisfies the tilted-plane condition, which gives rise to derivative discontinuities (see also~\cite{CapVigUll10,GoshenKraisler24}). 

However, approximate XC functionals may strongly violate the aforementioned exact conditions, leading, among other problems, to delocalization and static correlation errors. 
Incorporating known exact properties into an XC approximation during its development constraints its functional form, and can potentially lead to more accurate results for many systems. Recovering piecewise-linearity and the flat plane conditions have been demonstrated to lead to improved results. ~\cite{GouldDobson13,BajajKulik17,BajajKulik22,Prokopiu22,WodynskiArbuzKaupp21,burgess2023dft+} 
For example, achieving piecewise-linearity for an ensemble of different particle numbers via the use of a tuned range-separated hybrid functionals or by performing an ensemble generalization has been demonstrated to correct many of the ailments of standard DFT functionals, such as self-interaction error, orbital gap, the description of charge transfer excitations and the spurious fractional dissociation problem.\cite{Kronik_JCTC_review12, SteinKronikBaer_curvatures12,stein2009reliable,Stein10,KraislerKronik13, KraislerKronik14, KraislerSchmidt15, KraislerKronik15} 
Moreover, tuning range-separated double hybrids~\cite{Prokopiu22} has been used to enforce both piecewise-linearity and spin constancy, resulting in the minimization of both errors and improved dissociation curves, despite the presence of a non-physical maximum. 
In summary, the process of spin migration, namely the change of the system's spin $S$ at constant $N$, has been examined mainly from the perspective of the total energy.

In the present contribution, we focus on the behavior of the energy, the density, the orbitals, and the KS potential, while varying the ($z$-projection of the) spin of the system, $S$, accounting for both integer and fractional values.  
We make progress in the following directions: (i) As a first step towards a comprehensive understanding of spin migration processes, we consider various atomic systems, namely, H, He, Li, Be, B, C, Na, Mg, Si, and K, where transitions can occur between $s$-$s$, $s$-$p$ and $p$-$p$ orbitals. Three different types of behavior for the total energy have been identified.
(ii) We examine the changes versus $S$ both within and beyond the range $[-S_{\min},S_{\min}]$.
(iii) We explore the behavior of the frontier KS orbitals versus $S$ and (iv) monitor the changes in the KS potentials, $v_\KS^\s(\rr;S)$, for selected cases. In cases where the functional includes exact exchange, we discovered the emergence of plateaus in the potential when the spin reaches half-integer values; appearance of these plateaus corresponds to jumps in the frontier orbital energies. (v) Finally, the deviation of the spin-densities $n_\s(\rr)$ from piecewise-linearity is assessed and the relation to the energy deviation in the same cases is discussed.

The rest of the article is organized as follows: we summarize the relevant theoretical results in Sec.~\ref{Sec:theoretical_background}  and provide the numerical details of our calculations in Sec.~\ref{sec:numerical}. The results are presented in Sec.~\ref{Sec:results}, where we discuss them according to the total energy behavior: we describe the case of concave behavior in Sec.~\ref{sec:caseI}, a mixture of concave and convex behavior in Sec.~\ref{sec:caseII}, and the all concave case in Sec.~\ref{sec:caseIII}. Section~\ref{sec:conclusions} concludes the article. 
 
\section{Theoretical background}
\label{Sec:theoretical_background}

In this section, we outline the main exact features of any finite many-electron system, which emerge when varying its spin. These features serve as a reference point in analyzing approximate results. We are relying on previous studies, particularly Ref.~\cite{GoshenKraisler24}.

First, consider the case when the total number of electrons, $N$, is kept constant at an integer value, while the ($z$-projection of the) spin, $S$, varies within $[-S_\m, S_\m]$ (e.g., the spin varies between $-\thalf$ and $\thalf$ for Li or between $-\tfrac{3}{2}$ and $\tfrac{3}{2}$ for N). In the terminology of Ref.~\cite{GoshenKraisler24}, this is \emph{spin migration within the trapezoid} (more precisely,  on the trapeziod's side). Then, the ensemble ground state of the system consists of the states $\ket{\Psi_{N,-S_\m}}$, $\ket{\Psi_{N,-S_\m+1}}$, ..., $\ket{\Psi_{N,S_\m-1}}$ and $\ket{\Psi_{N,S_\m}}$, which all have the same total energy, and therefore we expect the ensemble total energy, $E(S)$, to be spin-independent. As a direct result, strictly within the range $(-S_\m, S_\m)$ the frontier orbitals must coincide and equal the negative of the ionization potential (IP): $\eps_\ho^\up = \eps_\ho^\dw = -I$.  In contrast, at and beyond the range boundary, the frontier orbital energies may change abruptly. This may happen due to the change in the states that participate in the ensemble, and the resultant slope of the energy, giving rise to the spin-migration derivative discontinuity predicted in Ref.~\cite{GoshenKraisler24,BurgessLinscottORegan24,CapVigUll10}. Plateaus in the KS potentials are expected to form as a result of this discontinuity~\cite{GoshenKraisler24}.

Outside the trapezoids, when $N$ is a constant integer, but $S > S_\m$ (e.g., Li with $S \in [\thalf,\tfrac{3}{2}]$), states with different values of $N$ and $S$ can contribute to the ensemble. Similar is the situation for $S < -S_\m$. Ref.~\cite{BurgessLinscottORegan24} studied this question in detail for selected atomic systems and established the fact that for high spin values the ensemble ground state may include contributions not only from those pure states that are closest to a given point in the $N_\up-N_\dw$ plane. However, in the systems we describe here, namely for Li, B, Na and Mg$^+$, 
atoms with $S \in [\thalf,\tfrac{3}{2}]$ and for He, Li$^+$, Be$^{2+}$, Be, Mg, C and Si  with $S \in [0,1]$, the only two contributing states are $\ket{\Psi_{N,S_\m}}$ and $\ket{\Psi_{N,S_\m+1}}$. We have verified this statement, relying on the existing tabulated experimental spectroscopic data from the National Institute of Standards (NIST)~\cite{NIST_lines}. The fact that our ensemble includes only the two aforementioned states means that the ensemble total energy is linear in the spin: 
\begin{equation} \label{eq:E_S__pl}
    E(S) = (S_\m+1-S) E(S_\m) + (S-S_\m) E(S_\m+1).
\end{equation}
To clearly illustrate the deviation of the energy obtained with approximate XC functionals, we refer to $\Delta E(S)$ -- the deviation of the energy $E(S)$ from the expected piecewise-linear (or constant) behavior. $\Delta E$ is nothing else but the difference between the left- and the right-hand sides of Eq.~\eqref{eq:E_S__pl}, both obtained within a given XC approximation.

The frontier orbitals are not equal anymore, but they are spin-independent. This can be shown relying on Janak's theorem~\cite{Janak78}, which identifies the KS eigenvalues with derivatives of the total energy with respect to the corresponding occupation numbers. Particularly, for the frontier orbitals, 
\begin{equation}
    \eps_\ho^\s = \frac{\de E}{\de f_\ho^\s} = \lp( \frac{\de E}{\de N_\s} \rp)_{N_\sbar},
\end{equation}
where we assumed that in each spin channel only the homo level can change its occupation. This is the scenario in all cases we consider in this work. Here $\sbar$ denotes the spin channel opposite to $\s$. Performing a change of variables from $N_\up$ and $N_\dw$ to $N= N_\up+N_\dw$ and $S=\thalf(N_\up-N_\dw)$, we find that
\begin{equation}
\lp( \frac{\de E}{\de N_\s} \rp)_{N_\sbar} = \lp( \frac{\de E}{\de N} \rp)_S \lp( \frac{\de N}{\de N_\s} \rp)_{N_\sbar} + \lp( \frac{\de E}{\de S} \rp)_N \lp( \frac{\de S}{\de N_\s} \rp)_{N_\sbar}.
\end{equation}
Using Eq.~\eqref{eq:E_S__pl}, we find that $(\de E/\de S)_N = E(S_\m+1) - E(S_\m) =: J_1$ -- the first spin-flip energy. Additionally, defining $\delta_\s$ to equal 1 for $\s=\up$ and -1 for $\s=\dw$, we arrive at the relation
\begin{equation}
    \eps_\ho^\s = \lp( \frac{\de E}{\de N} \rp)_S + \half \delta_\s J_1
\end{equation}
The value of the remaining unknown derivative depends on the third pure-state that contributes to the ensemble ground state at fractional electron number as it approaches $N$ from below. Different scenarios can occur in this regard, but one thing is common: this term does not depend on $S$. As a result, $\eps_\ho^\s$ is also $S$-independent, in the whole range $(S_\m, S_\m+1)$. As mentioned, the frontier orbital energies are not equal: their difference, $\eps_\ho^\up - \eps_\ho^\dw = J_1$ is the spin-flip energy.

The electron spin-densities $n_\s(\rr;S)$ are linear combinations of the pure-state densities that comprise the ensemble. In particular, in all ensembles that consist of two states (generally denoted here 0 and 1), the density must equal
\begin{equation} \label{eq:n_x}
    n^\s(\rr;x) = (1-x) n^\s_0(\rr) + x n^\s_1(\rr),
\end{equation}
where $x \in [0,1]$.
For example, for Li with $S \in [-\thalf,\thalf]$, 0 corresponds to $S=-\thalf$, 1 corresponds to $S=\thalf$ and $x = S+\thalf$. For Li with $S \in [\thalf,\tfrac{3}{2}]$, 0 corresponds to $S=\thalf$, 1 corresponds to $S=\tfrac{3}{2}$ and $x = S-\thalf$. To assess the deviation of a spin-density obtained in an approximate calculation from the expected behavior~\eqref{eq:n_x}, we first define
\begin{equation} \label{eq:d_x}
    d^\s(\rr;x) = n^\s(\rr;x) - \lp[ (1-x) n^\s_0(\rr) + x n^\s_1(\rr) \rp],
\end{equation}
which is the difference between the left- and the right-hand side of Eq.~\eqref{eq:n_x}, when all quantities are obtained within a given XC functional. For the exact case,  $d^\s(\rr;x)$ must equal zero for all $\rr$ and $x$. For approximations, this is not the case. Note, however, that $ d^\s(\rr;x)$ equals 0 for $x=0$ and 1, by construction, for any XC functional and that $\int  d^\s(\rr;x) \,\, d^3r =0$, as well%
~\footnote{In atomic calculations within the spherical approximation, it is common to plot not the density itself, but the quantity $4 \pi r^2 n(\rr)$. This quantity is easier to visualize and, among other advantages, reveals the shell structure of the density. The same applies also to $ d^\s(\rr;x)$ in the present context.}.
In addition, we define the quantity
\begin{equation} \label{eq:Q_x}
    Q^\s(x) = \int  \lp[ d^\s(\rr;x) \rp]^2 \,\, d^3r
\end{equation}
to better assess the deviation of spin-densities from piecewise-linearity. $Q^\s(x)$ depends only on the spin (via $x$), and not on $\rr$. For the exact case, $Q^\s(x)$ vanishes for all values of $x$, while for approximations it measures in a concise manner the degree of deviation from the expected behavior.


\section{Numerical details}
\label{sec:numerical}

All atomic calculations were carried out using the \texttt{ORCHID} code~\cite{KraislerMakovArgamanKelson09, KraislerMakovKelson10}, which performs electronic structure calculations of single atoms and ions, within a spherical approximation. We used a discrete logarithmic grid with 16000 points, on the interval $[r_\textrm{min}, r_\textrm{max}]$, where $r_\textrm{min} = e^{-13}/Z$ Bohr (with $Z$ being the atomic
number), and $r_\textrm{max} = 35$ Bohr.  For certain small systems (e.g., He) a smaller grid was used, to prevent numerical noise in cases where all relevant orbitals rapidly decay. 
Orbital-dependent XC functionals were treated within the Krieger-Li-Iafrate (KLI)~\cite{KLI92} approximation in the Optimized Effective Potential (OEP) formalism~\cite{KK08}. Note that in some cases when employing EXX, there is no convergence near the edges of the spin range, because one of the frontier orbitals becomes positive.

The variation of the spin $S$ was performed by simultaneously increasing $N_\up$ and decreasing $N_\dw$ in the calculations. This translated to the increase in the occupation number $f_{\ho,\up}$ in the Kohn-Sham system, and the simultaneous decrease in $f_{\ho,\dw}$.

\section{Results}
\label{Sec:results}

In this section, we present results for the atoms H, He, Li, Be, B, C, Na, Mg, Si and K  and some of their ions. We discuss in detail the dependence of the total energy, the frontier orbital energies, the KS potentials and the electron densities on the spin, $S$. Our findings can be subdivided into three cases, as detailed below. We start with the all-concave energy behavior, which is the most common case in the literature.

\subsection{CASE I: All-concave energy behavior}
\label{sec:caseI}
\begin{figure*}
    \centering
    \includegraphics[width=0.36\linewidth]{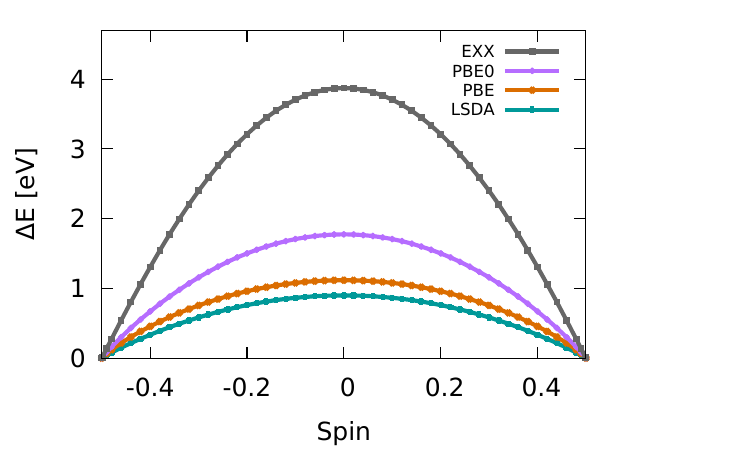}
    \hspace{-10mm}
   \includegraphics[width=0.36\linewidth]{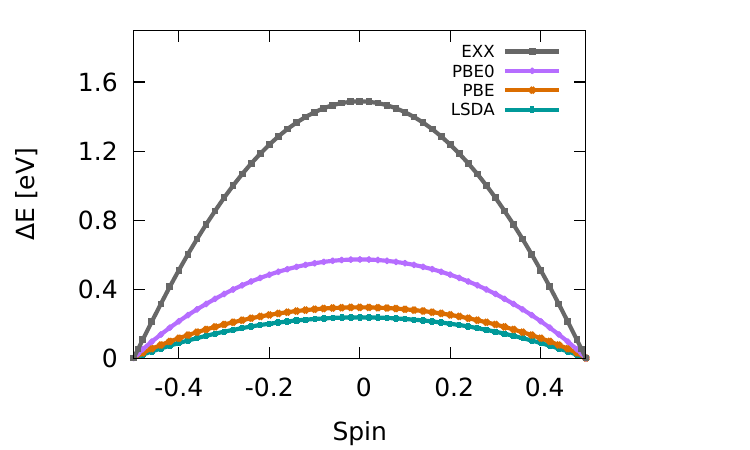}
   \hspace{-10mm}
    \includegraphics[width=0.36\linewidth]{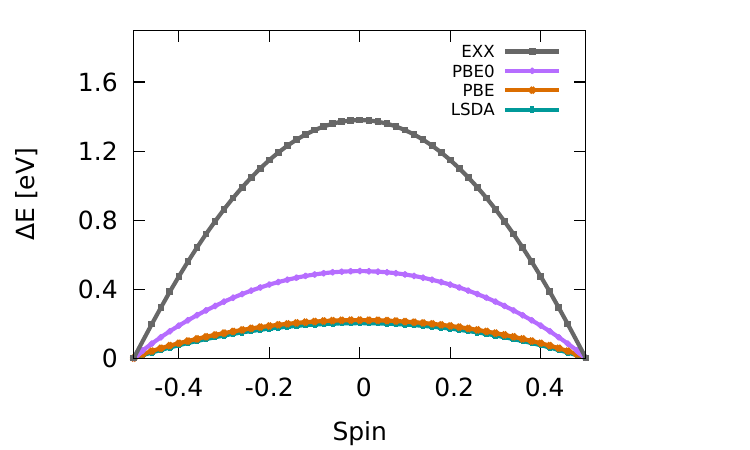}
    \caption{Deviation in energy, $\Delta E$, versus the fractional spin, $S$, for the H, Li and Na atoms (left to right), for different XC approximations (see Legend).} 
    \label{fig:Total_Energy_case_I}
\end{figure*}

Figure \ref{fig:Total_Energy_case_I} shows the deviation of the total energy, $\Delta E(S)$, with respect to  the fractional spin, $S$, for the H, Li, and Na systems, where the spin varies between $-0.5$ and $0.5$. In agreement with previous studies~\cite{CohenMoriSYang08,Prokopiu22}, all tested functionals, namely LSDA, PBE, PBE0, and EXX, overestimate the energy at fractional spins and result in a concave behavior for $\Delta E(S)$. The smallest deviation is observed for LSDA and PBE, and the largest deviation is observed for EXX. The deviation of PBE0 is between PBE and EXX, as expected, and approximately equals the linear combination of the former results, with weights of 0.75 and 0.25, respectively (see Table~S1 in the Suplementary Material (SM)).   While in all three atomic systems, the behavior of all functionals is similar, we observe a larger deviation for the H atom system, which reduces for Li and Na.

\begin{figure*}
    \centering
    \hspace{-12mm}
    \includegraphics[width=0.30\linewidth]{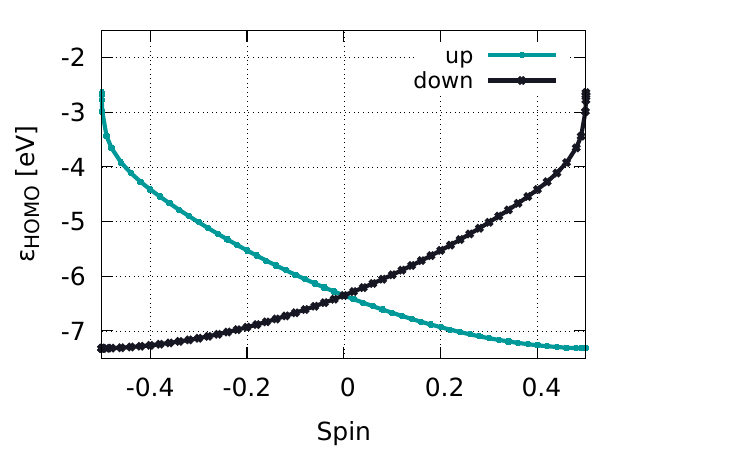}
    \hspace{-12mm}
   \includegraphics[width=0.30\linewidth]{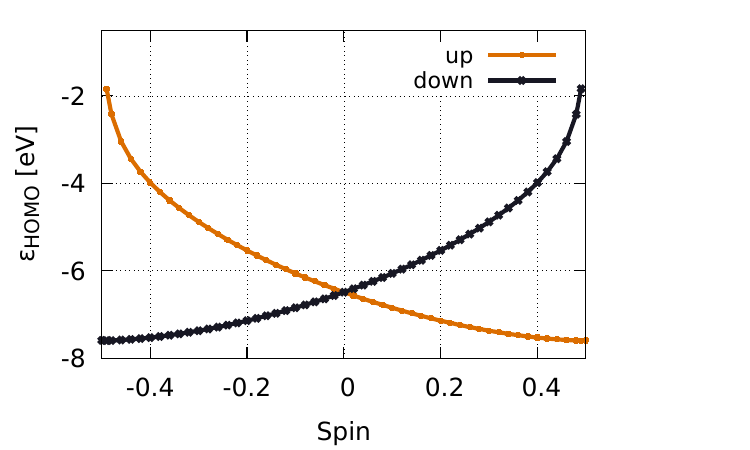}
   \hspace{-12mm}
    \includegraphics[width=0.30\linewidth]{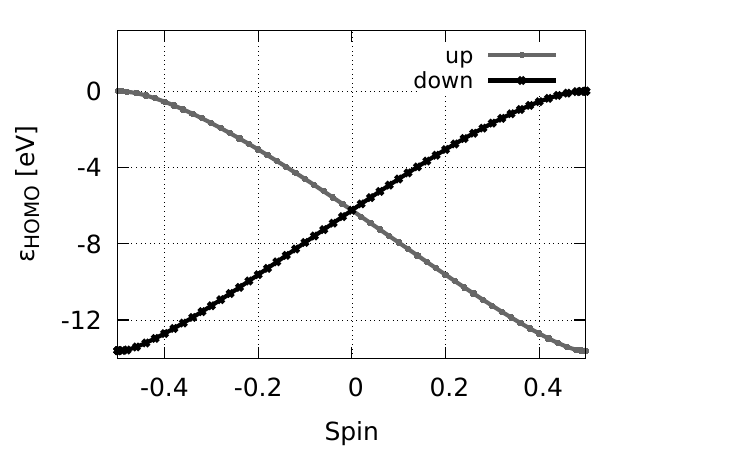}
    \hspace{-12mm}
    \includegraphics[width=0.30\linewidth]{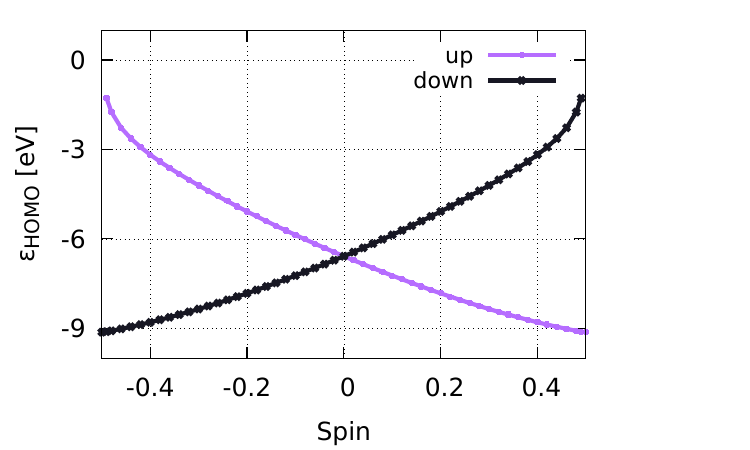}
    \hspace{-12mm}
    \caption{Frontier orbital energies for the H atom versus the fractional spin. From left to right: LSDA, PBE, EXX, PBE0 functionals. At the edges of the spin range,  the energy level of the unoccupied up/down channel is not shown in the plot. }
    \label{fig:H_HOMO}
\end{figure*}

To gain more insights into these systems, we next study the energy of the frontier orbitals with respect to fractional spins. For H, these are the $1s^\up$ and $1s^\dw$ orbitals, for Li -- $2s^\up$ and $2s^\dw$ and for Na -- $3s^\up$ and $3s^\dw$. The exact behavior should be a straight horizontal line with a possible jump in the frontier orbital value at the edges of the spin range. As expected from the behavior of the total energy, here again, we observe a deviation from the exact behavior in all three systems. Particularly, Fig.~\ref{fig:H_HOMO} demonstrates the behavior of the frontier orbitals for the H atom for the different functionals. The expected exact result is that both the up and the down orbital energies equal $-13.6$ eV, for all values of $S$. All functionals produce results that strongly deviate from this expectation: as one can see from the Figure, when increasing $S$, the energy value of the up channel decreases, and for the down channel the value increases. Even for EXX, which is expected to be exact for the H atom, we observe the above behavior. We attribute this deviation to the fact that while the Hartree and the EXX potentials cancel at the edges, i.e., at $S = \pm 0.5$, for any other value of $S$ they differ: $v_H(\rr) + v_{x,\s}^{exact}(\rr) = (1 - (0.5 + \delta_\s S)^2) v_H(\rr)$ (where $\delta_\up = 1$ and $\delta_\dw = -1$, as defined above)~\cite{GouldDobson13}.

\begin{figure*}
    \centering
    \hspace{-12mm}
    \includegraphics[width=0.30\linewidth]{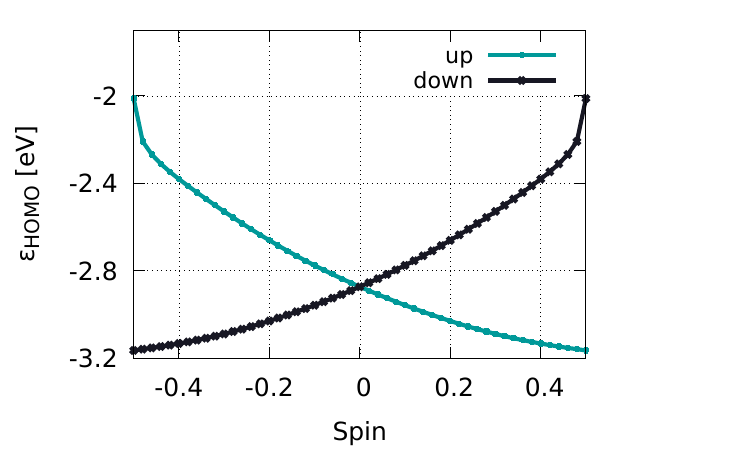}
    \hspace{-12mm}
   \includegraphics[width=0.30\linewidth]{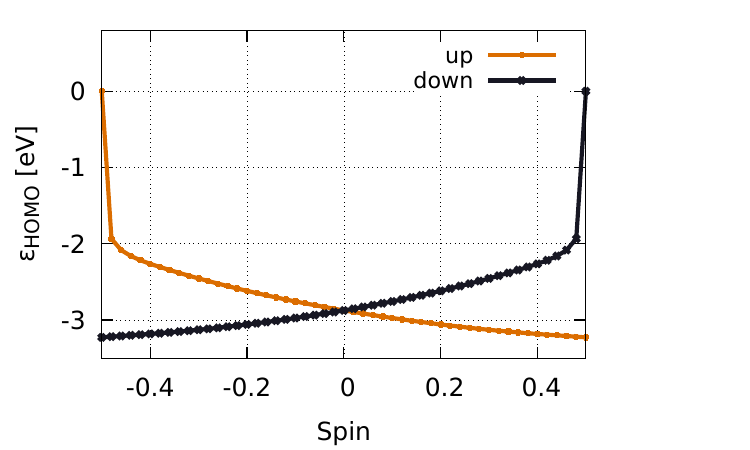}
    \hspace{-12mm}
    \includegraphics[width=0.30\linewidth]{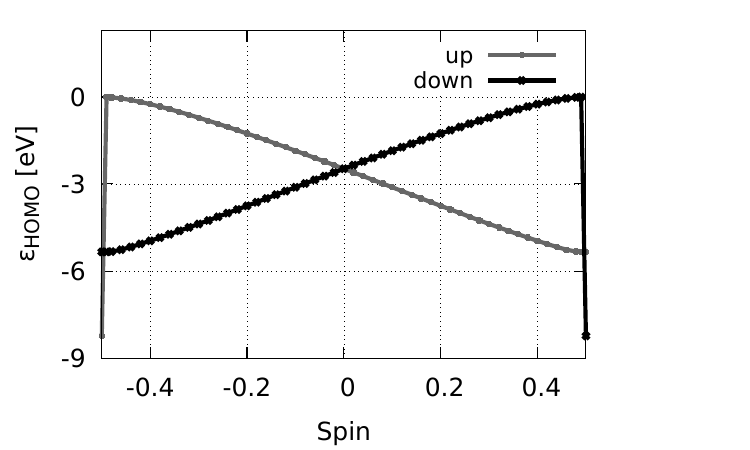}
    \hspace{-12mm}
    \includegraphics[width=0.30\linewidth]{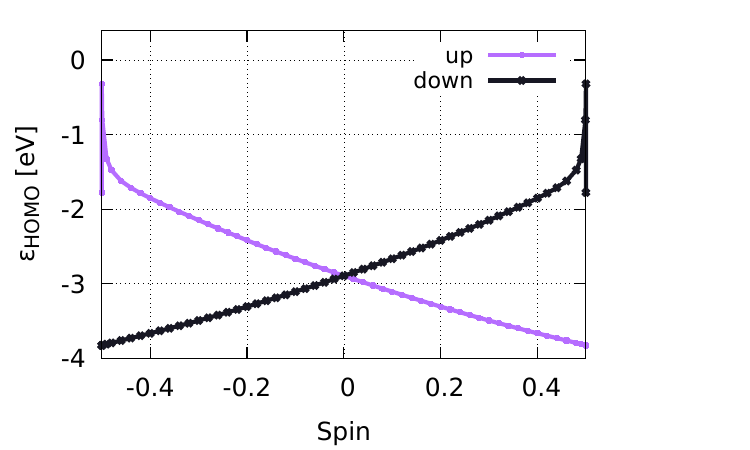}
    \hspace{-12mm}

    \centering
    \hspace{-12mm}
    \includegraphics[width=0.30\linewidth]{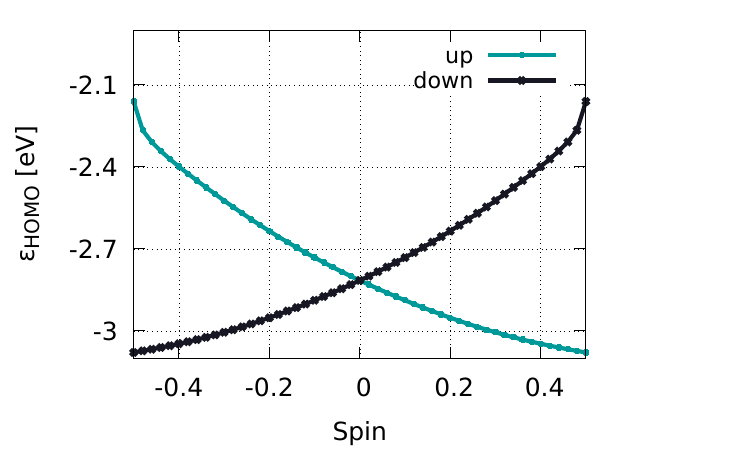}
    \hspace{-12mm}
   \includegraphics[width=0.30\linewidth]{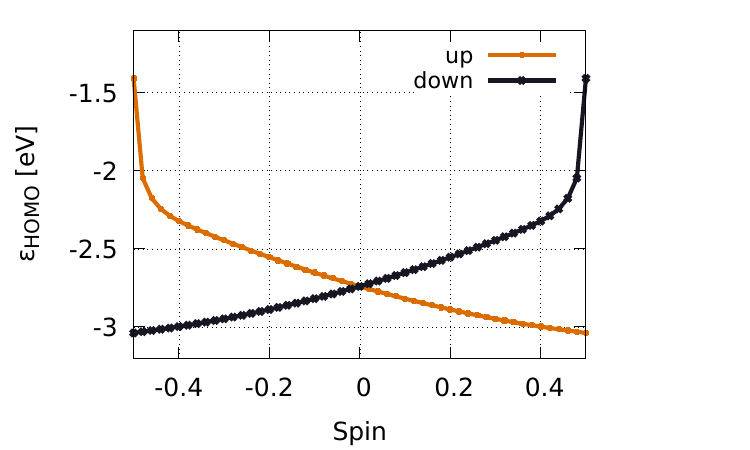}
    \hspace{-12mm}
    \includegraphics[width=0.30\linewidth]{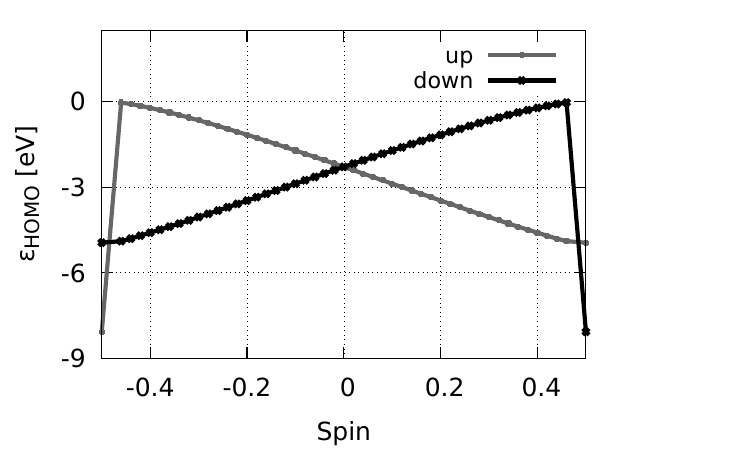}
    \hspace{-12mm}
    \includegraphics[width=0.30\linewidth]{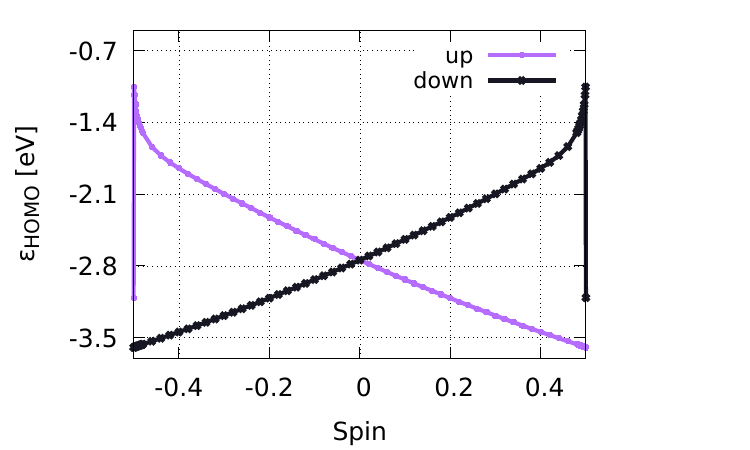}
    \hspace{-12mm}
    \caption{Frontier orbital energies for the Li atom (top) and the Na atom (bottom) versus the fractional spin. From left to right: LSDA, PBE, EXX, PBE0 functionals.}
   
    \label{fig:Li_Na_HOMO_GS}
\end{figure*}

Going beyond the one-electron system, we see additional interesting features for the different XC functionals, as shown in Fig.~\ref{fig:Li_Na_HOMO_GS}. For Li and Na, there is a similarity in the behavior of the frontier orbitals, as was observed for the total energies. 

For the LSDA functional, shown in Fig.~\ref{fig:Li_Na_HOMO_GS} (leftmost column), 
we observe a decrease in value of the up channel and a simultaneous increase of the down channel, similarly to the results for the H atom. As the total energies behave approximately parabolically, the resulting deviation in the frontier orbitals is approximately linear. However, the nonlinearity in the frontier orbital energies is apparent, which suggests that also in the total energies there are significant contributions beyond the parabolic term. The results for positive and negative values of $S$ are mirror images with the same value at $S=0$ as both levels coincide, as expected with identical occupation numbers for up and down channels which provide the unpolarized result. As for the case of H, we observe no jumps or sharp features.

In the case of the PBE functional shown in Fig.~\ref{fig:Li_Na_HOMO_GS} (second column from the left), the behavior is similar to that of the LSDA, with the difference that with the PBE functional, for both Li and Na, sharp features (`spikes') in the frontier orbitals near the edges ($S=-0.5$ or $0.5$) occur. 

The next functional we discuss is the EXX, which shows a very different behavior (Fig.~\ref{fig:Li_Na_HOMO_GS} (third column from the left).) 
Here, for both Li and Na systems, we observe a significant jump of the up frontier orbital energy at $S=-0.5^+$ and then a gradual \textbf{decrease} in this energy value as we increase $S$; similarly, we observe the same jump in the down frontier orbital energy at $S=0.5^-$ and then a decrease as we lower $S$. Unlike the feature of spikes we observed in PBE, the aforementioned jump in the EXX is instantaneous in $S$ and does not depend on the sampling of the $S$-axis. 

EXX is the only functional we studied, for which the unoccupied down level starts below the occupied up one; then, upon an infinitesimal spin change, a jump in the down level occurs, as expected. However, the jump is too high compared to the exact value, which results in a subsequent decrease in the frontier down energy level. The fact that we observed an unoccupied down KS level which lies below an occupied up one is not a forbidden scenario~\cite{KraislerMakovArgamanKelson09}: such an electronic configuration of the KS system is termed \emph{proper in a broad sense}. It is not uncommon in atomic systems calculated with LSDA or PBE~\cite{KraislerMakovKelson10}, and has been explicitly observed also for Li with EXX~\cite{Makmal09JCTC}.

For the PBE0 functional (Fig.~\ref{fig:Li_Na_HOMO_GS} (right column)), we experience the combination of the PBE spikes (upwards) and the jumps from EXX (downwards), which creates a numerical challenge. For this reason, we have also calculated the frontier orbitals for the LSDA0 case (shown in the SM, Fig.~S1). The frontier orbitals of LSDA0 also demonstrate an instantaneous jump at the edges of the $S$ range, which is smaller than the one observed for EXX, as only a fraction of the EXX is taken. Consequently, both frontier orbitals remain strictly bound for all values of $S$, which makes the numerical treatment easier.

\begin{figure}
    \centering
    \includegraphics[width=1.2\linewidth]{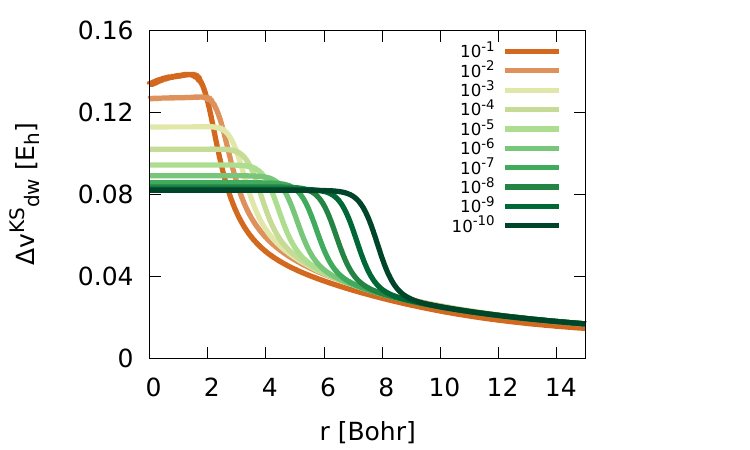}
    \caption{The difference between the $\dw$-KS potential for fractional $S=\thalf - \delta$ and the said potential at $S=\thalf$, obtained with the LSDA0 functional. See Legend for values of $\delta$.}
    \label{fig:potential for Li LDA0}
\end{figure}

The jumps in the frontier orbital energies necessitate the emergence of a plateau in the down-KS potential as we approach $S=0.5$ from below, similarly to the well-known case of fractional charges. Indeed, Fig.~\ref{fig:potential for Li LDA0} shows the difference of the KS potential of the fractional spin, $S$, from the KS potential at $S=0.5$. As we approach the values of $S=0.5$ from above, a clear spatial plateau is observed. It shares features known from fractional-charge plateaus: the plateau height converges to its asymptotic value as $1/\ln(0.5-S)$, and its width increases as $\ln(0.5-S)$ and it decays as $\sim 0.25/r$ (cf.\ Ref.~\cite{HodgsonKraisler17,Kraisler21}). To the best of our knowledge, plateaus in KS potentials are presented in this context for the first time.

Additional systems we have studied that exhibit the all-concave typical behavior are B and K, where the spin changes from $-\thalf$ to $\thalf$ and the transitions are between $s$ orbitals, Be and Mg atoms, where the spin changes from $S=0$ to $S=1$, and the transitions are between $s$ and $p$ orbitals. We also studied the C an Si atoms, where $S$ changes from 1 to 0, with transitions between the $p$ orbitals. The total energy curves with respect to the spin, in addition to the frontier orbitals, are shown in the SM, Figs.~S2 -- S6.

\begin{figure}
    \centering
    \includegraphics[width=1.2\linewidth]{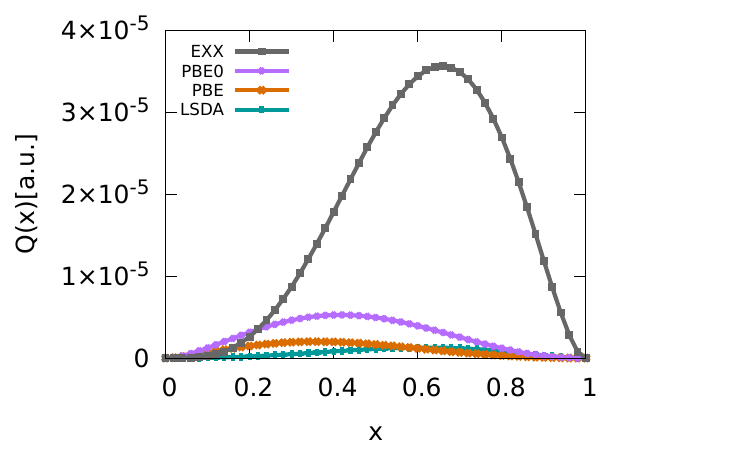}
    \caption{The quantity $Q^\dw(x)$ (defined in Eq.~\eqref{eq:Q_x}) for the Li atom, with various functionals (see Legend) where $x=S-\thalf$.
    }
    \label{fig:Li_Q}
\end{figure}

To obtain a complete picture, we next analyze the deviation of the electron density for fractional $S$, versus the expected ensemble density. We are using the quantity $Q^\dw(x)$, as defined in Eq.~\eqref{eq:Q_x}, and plot it for the case of the Li atom in Fig.~\ref{fig:Li_Q}, using various XC functionals ($Q^\up(x)$ is a mirror image of $Q^\dw(x)$, around $x=\thalf$). As shown in the Figure, the deviations of the down-density from piecewise-linearity are aligned with the deviations 
observed for the total energy (Fig.~\ref{fig:Total_Energy_case_I}), where the maximal deviation is observed for the EXX, followed by PBE0, and then  PBE and LSDA, which deviate the least. This might indicate that the errors in the total energies may be driven by errors from the density~\cite{kim2013understanding}.

\subsection{CASE II - mixed energy behavior}
\label{sec:caseII}

\begin{figure*}
    \centering
    \hspace{-10mm}
    \includegraphics[width=0.36\linewidth]{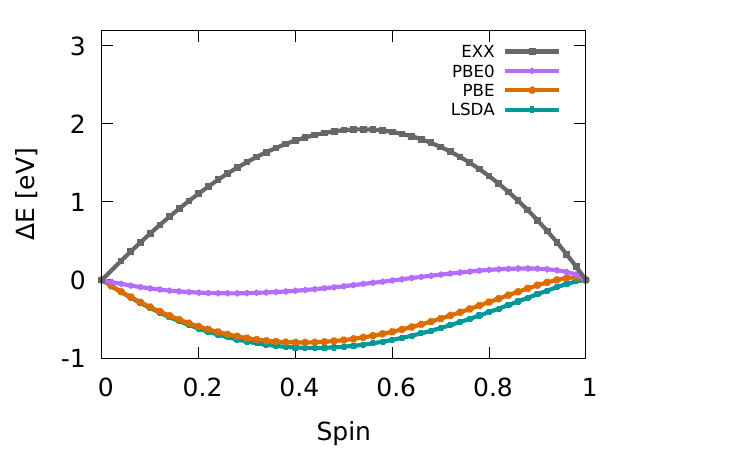}
    \hspace{-10mm}
   \includegraphics[width=0.36\linewidth]{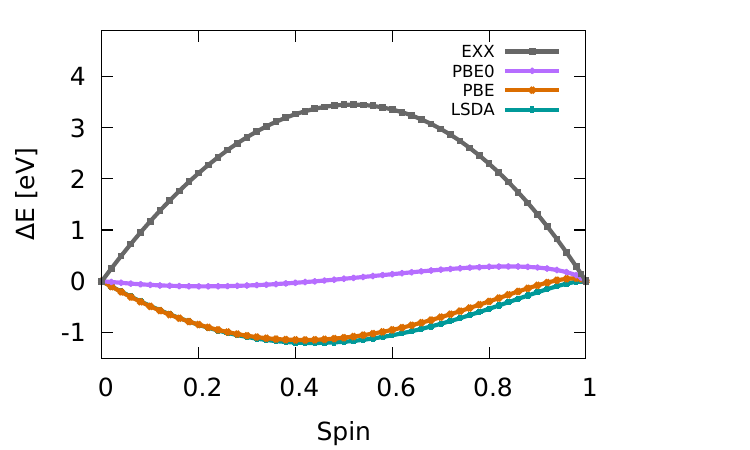}
    \hspace{-10mm}
    \includegraphics[width=0.36\linewidth]{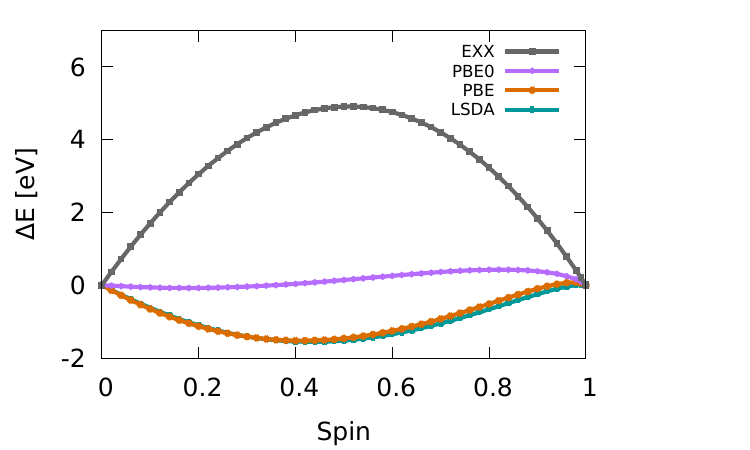}
    \hspace{-10mm}
    \caption{Deviation in energy, $\Delta E$, versus the fractional spin, $S$, for the He, Li$^+$ and Be$^{2+}$ systems (left to right), for different XC approximations (see Legend).}    
    \label{fig:He_like_systems}
\end{figure*}

The next systems under study are the He-like atoms, namely: He, Li$^+$, and Be$^{2+}$, where 2 electrons occupy the 1$s$ orbital and the spin value changes from $S=0$ to $S=1$, exceeding the range $[-S_\textrm{min}, S_\textrm{min}]$. Figure \ref{fig:He_like_systems} shows the behavior of the total energy with respect to fractional spin with the different functionals. 
Surprisingly, unlike the systems examined in Case I, 
here we see, for all three systems, an overestimation of the energy using the EXX functional and an underestimation of the energy for the PBE and LSDA functionals, resulting in a concave energy graph for the former and a convex one for the two latter. For the He-like atoms, the deviation of LSDA and PBE downwards is significantly smaller than the deviation of the EXX upwards. Contrary to systems of Case I, here the use of a global hybrid results in minimal deviation due to the opposite tendencies in the functionals. Notably, the deviation in the energy increases with an increase in the nuclear charge (here, unlike previous systems, the number of electrons is constant). 

\begin{figure*}
    \centering
    \hspace{-12mm}
    \includegraphics[width=0.30\linewidth]{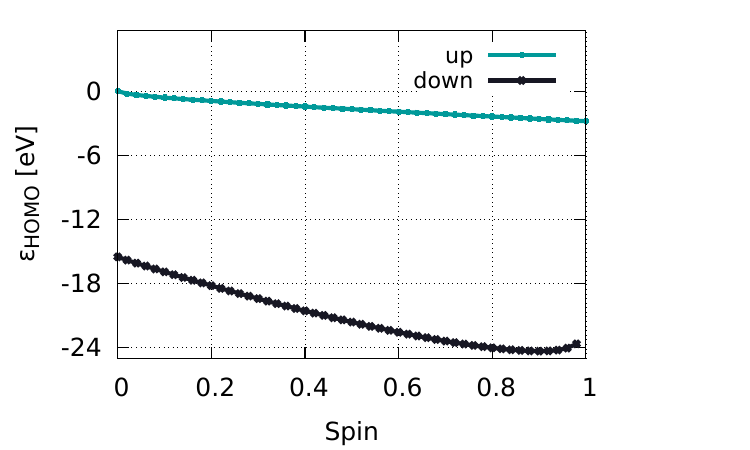}
    \hspace{-12mm}
   \includegraphics[width=0.30\linewidth]{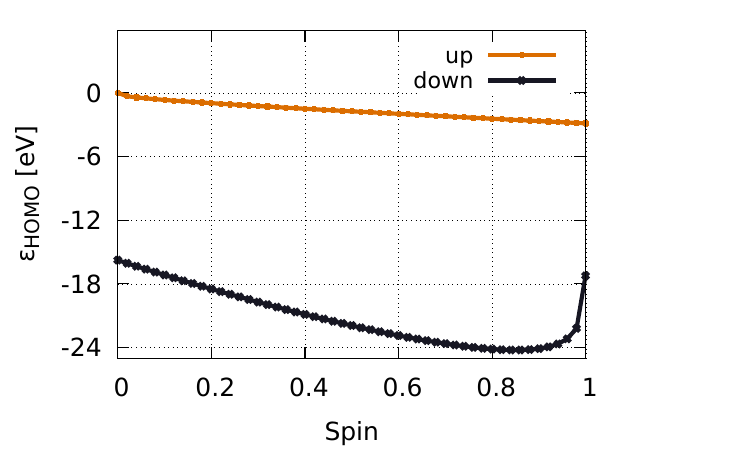}
    \hspace{-12mm}
    \includegraphics[width=0.30\linewidth]{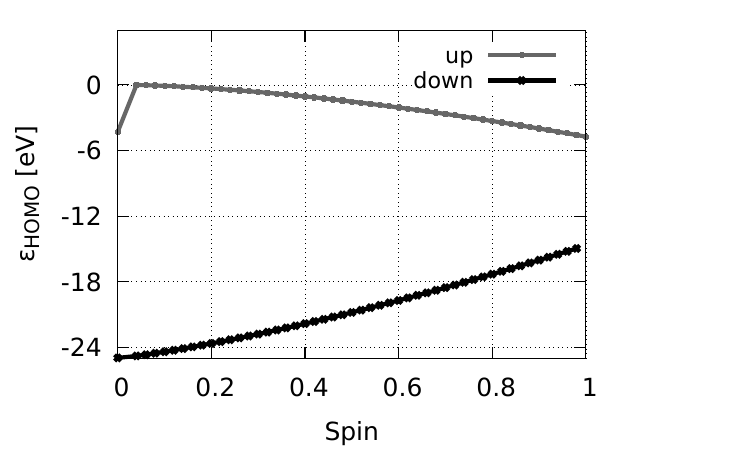}
    \hspace{-12mm}
    \includegraphics[width=0.30\linewidth]{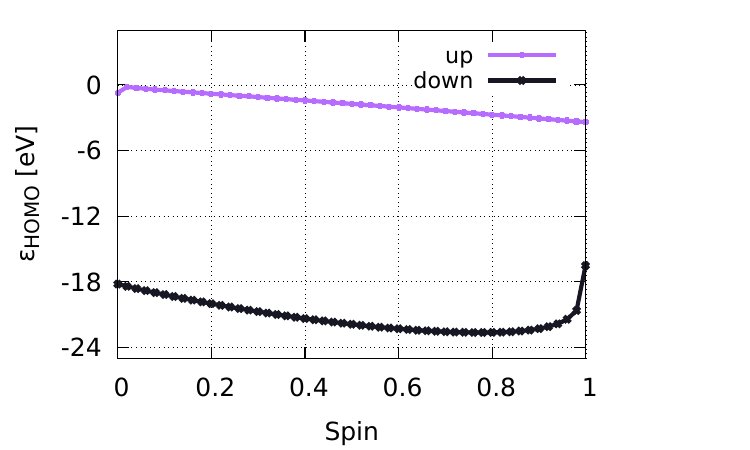}
    \hspace{-12mm}
    \caption{Frontier orbital energies for the He atom versus the fractional spin. From left to right: LSDA, PBE, EXX, PBE0 functionals. }
    \label{fig:He_HOMO}
\end{figure*}

The behavior of the frontier orbitals also shows similar trends for all the systems, as shown in the example of He in Fig.~\ref{fig:He_HOMO}. Frontier orbitals for Li$^+$ and Be$^{2+}$ are shown in Figs.~S7 and~S8, respectively. As can be seen from Fig.~\ref{fig:He_HOMO}, deviations of the exact behavior (namely, a straight horizontal line) are observed: for the LSDA functional emptying the down channel results in a decrease in the orbital energy, and as the up channel is gradually occupied, we observe a decrease in this channel as well. As before, the changes in the frontier orbitals are approximately linear; their slope is dictated by the parabolic term of the energy.

The frontier orbitals obtained from the PBE functional exhibit a similar behavior, with an approximately linear decrease in the energy levels. At $S=1$ we see spikes in the orbital energies, which are reminiscent of the spikes observed in PBE for H, Li, and Na, and seem to be a characteristic feature of the PBE functional. In the case of the EXX functional, in Fig.~\ref{fig:He_HOMO} (third from the left) we observe an increase in the down channel energy, as we gradually remove a down electron to occupy the up channel. At $S=0$ we observe a jump in the orbital energy, followed by a decrease as we progress towards $S=1$. As expected, PBE0 (Fig.~\ref{fig:He_HOMO} (right)) shows a combination of the features observed in PBE and EXX: we observe a medium-height spike at $S=1$ and a small jump in the up channel at $S=0$.

Interestingly, the behavior of the He-like systems is quite different than the one observed for Be and Mg (described in Sec.~\ref{sec:caseI}) despite the similarity in the electronic structure. 
Such a behavior is also obtained for the Li atom, where the spin changes from $S=0.5$ to $S=1.5$, as shown in the SM, Figs.~S9 and~S10.
We note that in all the systems presented here, we empty one of the spin channels. 

\begin{figure}
    \centering
    \mbox{
    \includegraphics[width=0.59\linewidth]{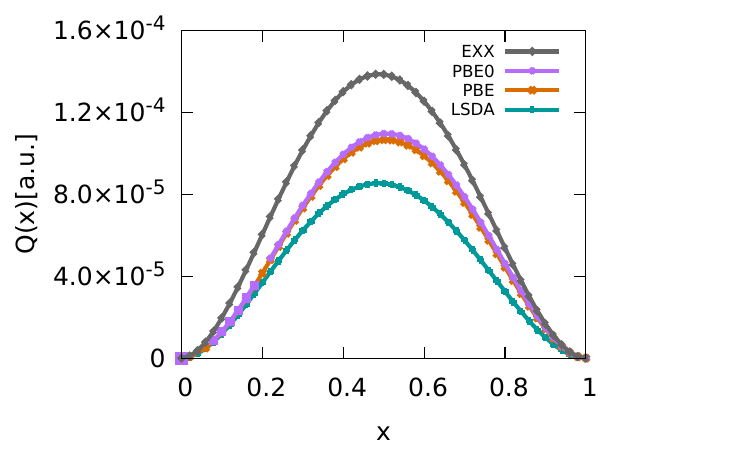}
    \hspace{-10mm}
    \includegraphics[width=0.59\linewidth]{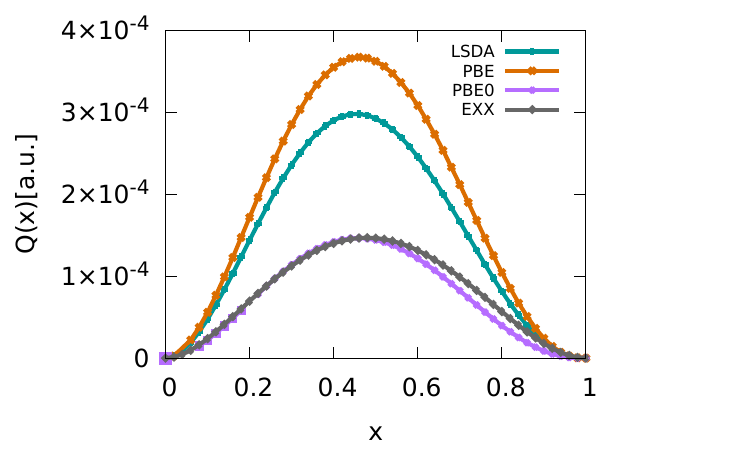}
    }
    \caption{The quantities $Q^\up(x)$ (left) and $Q^\dw(x)$ (right) for the He atom, with various functionals (see Legend), where $x=S$. }
    \label{fig:Q_He}
\end{figure}

Finally, analyzing the deviation of the density from the expected piecewise-linear behavior using the $Q^\s(x)$ measure is shown for the case of the He atom in Fig.~\ref{fig:Q_He}. Interestingly, for $Q^\dw(x)$ the maximum deviation from the ensemble density is observed for PBE, followed by LSDA, and the minimal deviation is obtained for EXX and PBE0. For $Q^\up(x)$, EXX deviates the most, followed by PBE and PBE0, while LSDA is the most accurate. All deviations $Q^\up(x)$ are lower than those of $Q^\dw(x)$. The errors in the density are not in line with the errors observed for the energy (cf.\ Ref.~\cite{kim2013understanding}). 

\subsection{CASE III - All convex energy behavior}
\label{sec:caseIII}

\begin{figure*}
    \centering
    \hspace{-10mm}
    \includegraphics[width=0.52\linewidth]{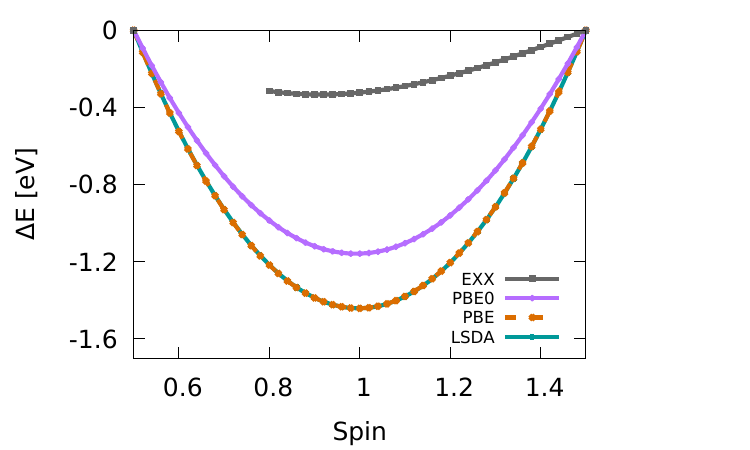}
    \hspace{-10mm}
   \includegraphics[width=0.52\linewidth]{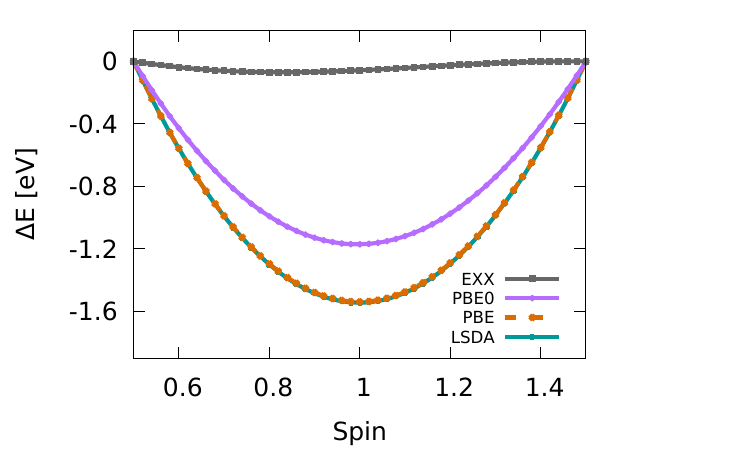}
   \hspace{-10mm}
    \caption{Deviation in energy, $\Delta E$, versus the fractional spin, $S$, for Na and  Mg$^+$ (left to right), for different XC approximations (see Legend). }
    \label{fig:Case_III_Total_Energy_Ex}
\end{figure*}

For the Na and Mg{$^+$} systems, where the spin changes from $S=0.5$ to $S=1.5$ and $S$ exceeds the range $[-S_{\min},S_{\min}]$ we observe a third type of behavior. 
While these systems are similar in electronic structure to the case of Li, the behavior is quite different, which indicates that it is very challenging to predict in advance the type of deviation of different XC functionals. 
Figure \ref{fig:Case_III_Total_Energy_Ex} shows the total energy with respect to fractional spin, in the range of $S=0.5$ to $S=1.5$, for the aforementioned atomic systems. Unlike previous cases, we observe an underestimation of the total energy in all tested functionals, very different from the 'typical' Case I deviation, reported in Sec.~\ref{sec:caseI}. Additionally, unlike previous cases, here PBE and LSDA functionals deviate the most, while EXX exhibits the smallest deviation (with a particularly small deviation shown for the Mg$^+$ atom), 
and consequently, the PBE0 errors are in between PBE and EXX.

\begin{figure*}
    \centering
    \hspace{-12mm}
    \includegraphics[width=0.30\linewidth]{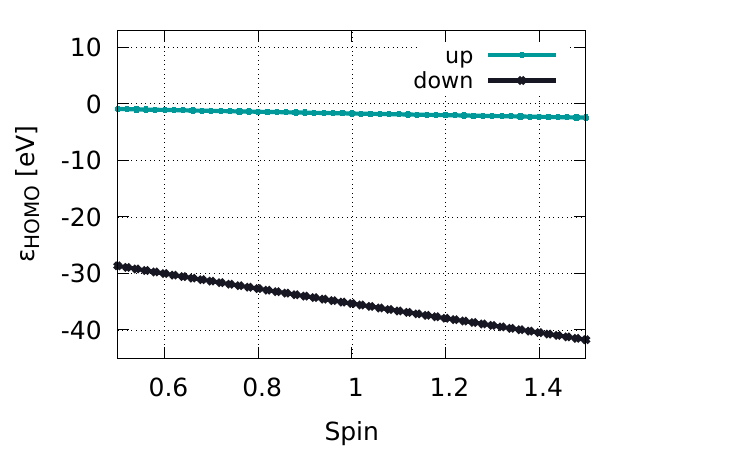}
    \hspace{-12mm}
   \includegraphics[width=0.30\linewidth]{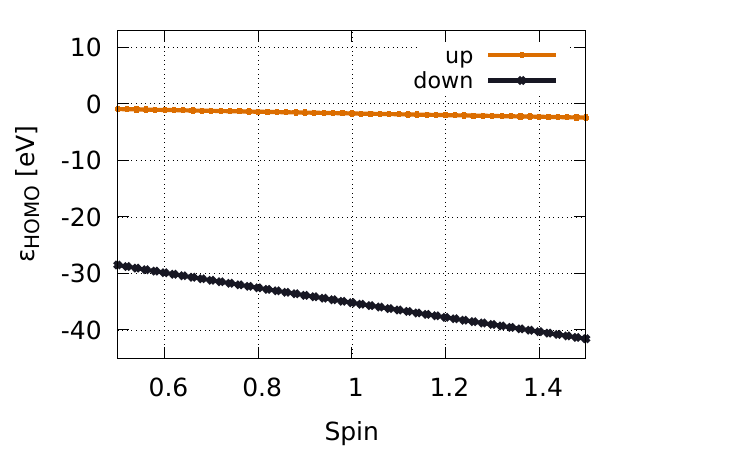}
    \hspace{-12mm}
    \includegraphics[width=0.30\linewidth]{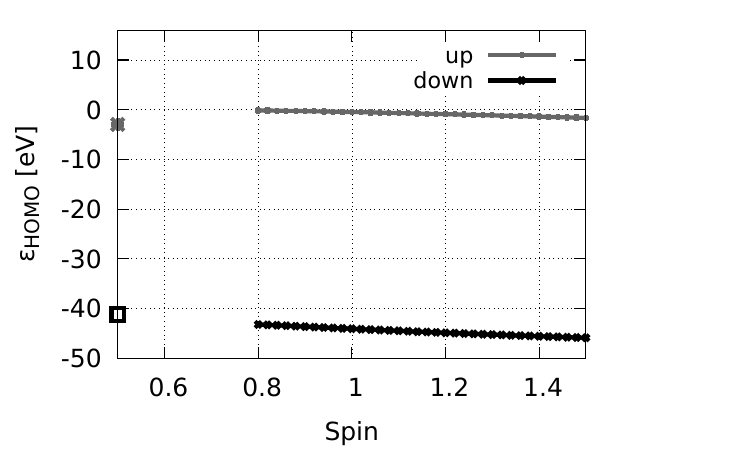}
    \hspace{-12mm}
    \includegraphics[width=0.30\linewidth]{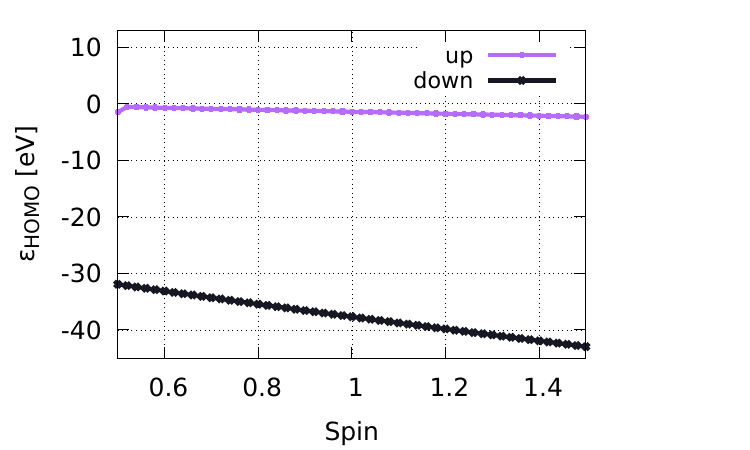}
    \hspace{-12mm}
    \caption{Frontier orbital energies for the Na atom versus the fractional spin, for different XC functionals: LSDA, PBE, EXX and PBE0 (left to right).}
    \label{fig:Na_EX_HOMO}
\end{figure*}

Figure \ref{fig:Na_EX_HOMO} shows the behavior of the frontier orbitals of Na, and Figure~S11 in the SM shows the frontier orbitals for Mg{$^+$}. As we empty the down channel, we observe a large decrease in the down orbital energy, in addition to a  (much smaller) decrease in the up channel energy for all functionals. For the EXX functional, we experience difficultes with convergence for $0.5 \leqs S \leqs 0.8$, as the up frontier orbital becomes unbound. Still, since  
the orbital energy value at $S=0.5$ is lower than the value at  $S=0.8$, 
we expect a jump in up the orbital energy for EXX. PBE0  results (Fig.~\ref{fig:Na_EX_HOMO} (right)), which converged in the whole $S$ range, indeed show a minute jump in the up channel at $S = 0.5^+$.

\begin{figure}
    \centering     
    \mbox{
    \includegraphics[width=0.59\linewidth]{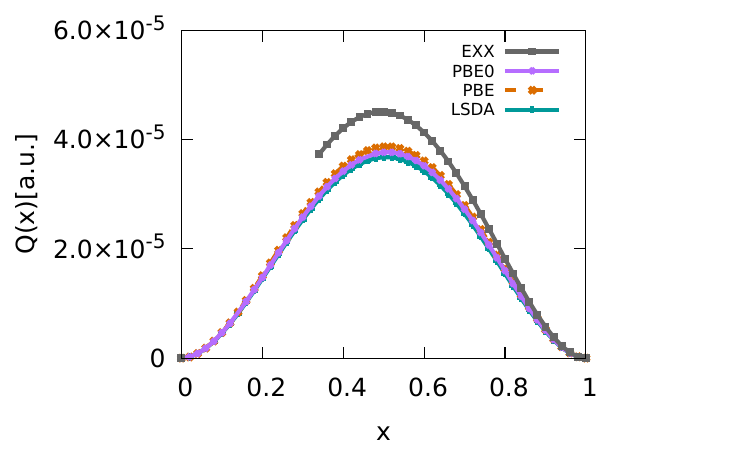}
    \hspace{-10mm}
    \includegraphics[width=0.59\linewidth]{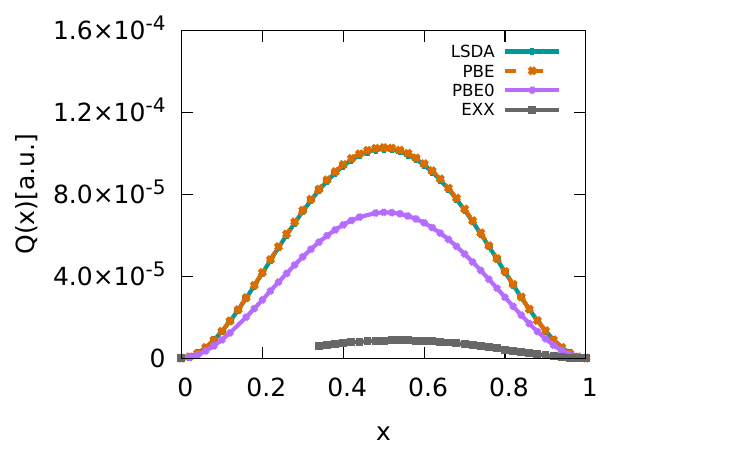}
    }
    \caption{The quantities $Q^\up(x)$ (left) and $Q^\dw(x)$ (right) for the Na atom, with various functionals (see Legend) where $x=S + 0.5$. }

    \label{fig:Q_Na}
\end{figure}

The deviation from the expected ensemble density for Na, using the $Q^\s(x)$ measure, is shown in Fig.~\ref{fig:Q_Na}. For $Q^\dw(x)$, the maximal, almost identical, deviation is observed for the LSDA and PBE functionals, and a minimal deviation for the EXX functional, while results for PBE0 are in between. The errors are compatible with the errors observed in the total energy, as in Case I. In contrast, for $Q^\up(x)$, EXX deviates the most, while all the other functionals yield a similar and lower result. 

\section{Conclusions}  \label{sec:conclusions}
In this manuscript, we have analyzed the deviation of the total energy, the frontier orbitals, the potential and the density versus the spin, for atomic systems, using different XC functionals. For the systems studied here, we identified three different types of behavior. Comparing to the known trend, where the XC functionals overestimate the total energy and provide concave curves with respect to fractional spins, we revealed a richer picture, where in some atomic systems all the XC functionals consistently yield a concave behavior, in some systems a consistent convex behavior, and yet several atoms, where we observe both concave and convex deviations, depending on the XC functional.
We demonstrate that in all cases the approximately parabolic deviation in the total energy results in approximately linear behavior of the frontier orbitals. We showed that the EXX, PBE0, and LSDA0 functionals exhibit a jump in the orbital energies, which is manifested as a plateau in the corresponding KS potential.
We additionally tested the deviation of the calculated approximate densities from the expected ensemble density and showed that in some cases the deviations are compatible, and in some they are not compatible with those of the total energy. This indicates that multiple sources of error may lead to these deviations.

As the errors exhibited by functionals for fractional spins are related to the description of static correlation, which is important in many chemical systems, the need to obtain a correct behavior versus fractional spin is highly important.
Insights provided in this work in analyzing errors and different types of deviations is instrumental in paving the way to their future correction.  Correcting the behavior of existing XC functionals and reproducing exact known conditions can lead to improvement in the prediction power of different properties with DFT, as has been demonstrated previously for the case of fractional charges.

\section*{Supplementary Material}
Supporting calculations for additional atomic systems are presented, according to the three cases described in the main text.

\acknowledgments{We acknowledge fruitful discussions with Yuli Goshen.
T.\ S.\ acknowledges support from the  European Research Council (ERC) Grant No. 101115429, AstroMol.}

\newpage

\mbox{}

\newpage 

\mbox{}


\bibliography{bib_EK_2023.bib}

\begin{thebibliography}{52}%
\makeatletter
\providecommand \@ifxundefined [1]{%
 \@ifx{#1\undefined}
}%
\providecommand \@ifnum [1]{%
 \ifnum #1\expandafter \@firstoftwo
 \else \expandafter \@secondoftwo
 \fi
}%
\providecommand \@ifx [1]{%
 \ifx #1\expandafter \@firstoftwo
 \else \expandafter \@secondoftwo
 \fi
}%
\providecommand \natexlab [1]{#1}%
\providecommand \enquote  [1]{``#1''}%
\providecommand \bibnamefont  [1]{#1}%
\providecommand \bibfnamefont [1]{#1}%
\providecommand \citenamefont [1]{#1}%
\providecommand \href@noop [0]{\@secondoftwo}%
\providecommand \href [0]{\begingroup \@sanitize@url \@href}%
\providecommand \@href[1]{\@@startlink{#1}\@@href}%
\providecommand \@@href[1]{\endgroup#1\@@endlink}%
\providecommand \@sanitize@url [0]{\catcode `\\12\catcode `\$12\catcode `\&12\catcode `\#12\catcode `\^12\catcode `\_12\catcode `\%12\relax}%
\providecommand \@@startlink[1]{}%
\providecommand \@@endlink[0]{}%
\providecommand \url  [0]{\begingroup\@sanitize@url \@url }%
\providecommand \@url [1]{\endgroup\@href {#1}{\urlprefix }}%
\providecommand \urlprefix  [0]{URL }%
\providecommand \Eprint [0]{\href }%
\providecommand \doibase [0]{https://doi.org/}%
\providecommand \selectlanguage [0]{\@gobble}%
\providecommand \bibinfo  [0]{\@secondoftwo}%
\providecommand \bibfield  [0]{\@secondoftwo}%
\providecommand \translation [1]{[#1]}%
\providecommand \BibitemOpen [0]{}%
\providecommand \bibitemStop [0]{}%
\providecommand \bibitemNoStop [0]{.\EOS\space}%
\providecommand \EOS [0]{\spacefactor3000\relax}%
\providecommand \BibitemShut  [1]{\csname bibitem#1\endcsname}%
\let\auto@bib@innerbib\@empty
\bibitem [{\citenamefont {Kohn}\ and\ \citenamefont {Sham}(1965)}]{KS65}%
  \BibitemOpen
  \bibfield  {author} {\bibinfo {author} {\bibfnamefont {W.}~\bibnamefont {Kohn}}\ and\ \bibinfo {author} {\bibfnamefont {L.~J.}\ \bibnamefont {Sham}},\ }\href@noop {} {\bibfield  {journal} {\bibinfo  {journal} {Phys. Rev.}\ }\textbf {\bibinfo {volume} {140}},\ \bibinfo {pages} {A1133} (\bibinfo {year} {1965})}\BibitemShut {NoStop}%
\bibitem [{\citenamefont {Parr}\ and\ \citenamefont {Yang}(1989)}]{PY}%
  \BibitemOpen
  \bibfield  {author} {\bibinfo {author} {\bibfnamefont {R.~G.}\ \bibnamefont {Parr}}\ and\ \bibinfo {author} {\bibfnamefont {W.}~\bibnamefont {Yang}},\ }\href@noop {} {\emph {\bibinfo {title} {{Density-Functional Theory of Atoms and Molecules}}}}\ (\bibinfo  {publisher} {Oxford University Press},\ \bibinfo {year} {1989})\BibitemShut {NoStop}%
\bibitem [{\citenamefont {Maurer}\ \emph {et~al.}(2019)\citenamefont {Maurer}, \citenamefont {Freysoldt}, \citenamefont {Reilly}, \citenamefont {Brandenburg}, \citenamefont {Hofmann}, \citenamefont {Bj{\"{o}}rkman}, \citenamefont {Leb{\`{e}}gue},\ and\ \citenamefont {Tkatchenko}}]{Maurer19}%
  \BibitemOpen
  \bibfield  {author} {\bibinfo {author} {\bibfnamefont {R.~J.}\ \bibnamefont {Maurer}}, \bibinfo {author} {\bibfnamefont {C.}~\bibnamefont {Freysoldt}}, \bibinfo {author} {\bibfnamefont {A.~M.}\ \bibnamefont {Reilly}}, \bibinfo {author} {\bibfnamefont {J.~G.}\ \bibnamefont {Brandenburg}}, \bibinfo {author} {\bibfnamefont {O.~T.}\ \bibnamefont {Hofmann}}, \bibinfo {author} {\bibfnamefont {T.}~\bibnamefont {Bj{\"{o}}rkman}}, \bibinfo {author} {\bibfnamefont {S.}~\bibnamefont {Leb{\`{e}}gue}},\ and\ \bibinfo {author} {\bibfnamefont {A.}~\bibnamefont {Tkatchenko}},\ }\href@noop {} {\bibfield  {journal} {\bibinfo  {journal} {Annu. Rev. Mater. Res.}\ }\textbf {\bibinfo {volume} {49}},\ \bibinfo {pages} {3.1} (\bibinfo {year} {2019})}\BibitemShut {NoStop}%
\bibitem [{\citenamefont {Tsuneda}(2014)}]{tsuneda2014density}%
  \BibitemOpen
  \bibfield  {author} {\bibinfo {author} {\bibfnamefont {T.}~\bibnamefont {Tsuneda}},\ }\href@noop {} {\emph {\bibinfo {title} {Density functional theory in quantum chemistry}}}\ (\bibinfo  {publisher} {Springer},\ \bibinfo {year} {2014})\BibitemShut {NoStop}%
\bibitem [{\citenamefont {Giustino}(2014)}]{Giustino14_materials}%
  \BibitemOpen
  \bibfield  {author} {\bibinfo {author} {\bibfnamefont {F.}~\bibnamefont {Giustino}},\ }\href@noop {} {\emph {\bibinfo {title} {{Materials modelling using density functional theory: properties and predictions}}}}\ (\bibinfo  {publisher} {Oxford University Press},\ \bibinfo {year} {2014})\BibitemShut {NoStop}%
\bibitem [{\citenamefont {Sholl}\ and\ \citenamefont {Steckel}(2011)}]{ShollSteckel11}%
  \BibitemOpen
  \bibfield  {author} {\bibinfo {author} {\bibfnamefont {D.}~\bibnamefont {Sholl}}\ and\ \bibinfo {author} {\bibfnamefont {J.}~\bibnamefont {Steckel}},\ }\href@noop {} {\emph {\bibinfo {title} {{Density Functional Theory: A Practical Introduction}}}}\ (\bibinfo  {publisher} {Wiley},\ \bibinfo {year} {2011})\BibitemShut {NoStop}%
\bibitem [{\citenamefont {Huang}\ and\ \citenamefont {Carter}(2008)}]{huang2008advances}%
  \BibitemOpen
  \bibfield  {author} {\bibinfo {author} {\bibfnamefont {P.}~\bibnamefont {Huang}}\ and\ \bibinfo {author} {\bibfnamefont {E.~A.}\ \bibnamefont {Carter}},\ }\href@noop {} {\bibfield  {journal} {\bibinfo  {journal} {Annu. Rev. Phys. Chem.}\ }\textbf {\bibinfo {volume} {59}},\ \bibinfo {pages} {261} (\bibinfo {year} {2008})}\BibitemShut {NoStop}%
\bibitem [{\citenamefont {Cramer}(2004)}]{Cramer04}%
  \BibitemOpen
  \bibfield  {author} {\bibinfo {author} {\bibfnamefont {C.~J.}\ \bibnamefont {Cramer}},\ }\href@noop {} {\emph {\bibinfo {title} {{Essentials of Computational Chemistry: Theories and Models}}}}\ (\bibinfo  {publisher} {Wiley},\ \bibinfo {year} {2004})\BibitemShut {NoStop}%
\bibitem [{\citenamefont {Perdew}\ and\ \citenamefont {Schmidt}(2001)}]{PerdewSchmidt01}%
  \BibitemOpen
  \bibfield  {author} {\bibinfo {author} {\bibfnamefont {J.~P.}\ \bibnamefont {Perdew}}\ and\ \bibinfo {author} {\bibfnamefont {K.}~\bibnamefont {Schmidt}},\ }in\ \href@noop {} {\emph {\bibinfo {booktitle} {Density Functional Theory and Its Application to Materials}}},\ \bibinfo {editor} {edited by\ \bibinfo {editor} {\bibfnamefont {V.}~\bibnamefont {{Van Doren}}}, \bibinfo {editor} {\bibfnamefont {C.}~\bibnamefont {{Van Alsenoy}}},\ and\ \bibinfo {editor} {\bibfnamefont {P.}~\bibnamefont {Geerlings}}}\ (\bibinfo  {publisher} {AIP},\ \bibinfo {year} {2001})\BibitemShut {NoStop}%
\bibitem [{\citenamefont {Perdew}\ \emph {et~al.}(2009)\citenamefont {Perdew}, \citenamefont {Ruzsinszky}, \citenamefont {Constantin}, \citenamefont {Sun},\ and\ \citenamefont {Csonka}}]{Perdew09_perplexed}%
  \BibitemOpen
  \bibfield  {author} {\bibinfo {author} {\bibfnamefont {J.~P.}\ \bibnamefont {Perdew}}, \bibinfo {author} {\bibfnamefont {A.}~\bibnamefont {Ruzsinszky}}, \bibinfo {author} {\bibfnamefont {L.~A.}\ \bibnamefont {Constantin}}, \bibinfo {author} {\bibfnamefont {J.}~\bibnamefont {Sun}},\ and\ \bibinfo {author} {\bibfnamefont {G.~I.}\ \bibnamefont {Csonka}},\ }\href@noop {} {\bibfield  {journal} {\bibinfo  {journal} {J. Chem. Theory Comp.}\ }\textbf {\bibinfo {volume} {5}},\ \bibinfo {pages} {902} (\bibinfo {year} {2009})}\BibitemShut {NoStop}%
\bibitem [{\citenamefont {Kaplan}\ \emph {et~al.}(2023)\citenamefont {Kaplan}, \citenamefont {Levy},\ and\ \citenamefont {Perdew}}]{KaplanLevyPerdew23}%
  \BibitemOpen
  \bibfield  {author} {\bibinfo {author} {\bibfnamefont {A.~D.}\ \bibnamefont {Kaplan}}, \bibinfo {author} {\bibfnamefont {M.}~\bibnamefont {Levy}},\ and\ \bibinfo {author} {\bibfnamefont {J.~P.}\ \bibnamefont {Perdew}},\ }\href@noop {} {\bibfield  {journal} {\bibinfo  {journal} {Annu. Rev. Phys. Chem.}\ }\textbf {\bibinfo {volume} {74}},\ \bibinfo {pages} {193} (\bibinfo {year} {2023})}\BibitemShut {NoStop}%
\bibitem [{\citenamefont {Sun}\ \emph {et~al.}(2015)\citenamefont {Sun}, \citenamefont {Ruzsinszky},\ and\ \citenamefont {Perdew}}]{Sun_SCAN_PRL15}%
  \BibitemOpen
  \bibfield  {author} {\bibinfo {author} {\bibfnamefont {J.}~\bibnamefont {Sun}}, \bibinfo {author} {\bibfnamefont {A.}~\bibnamefont {Ruzsinszky}},\ and\ \bibinfo {author} {\bibfnamefont {J.~P.}\ \bibnamefont {Perdew}},\ }\href@noop {} {\bibfield  {journal} {\bibinfo  {journal} {Phys. Rev. Lett.}\ }\textbf {\bibinfo {volume} {115}},\ \bibinfo {pages} {036402} (\bibinfo {year} {2015})}\BibitemShut {NoStop}%
\bibitem [{\citenamefont {Cuevas-Saavedra}\ \emph {et~al.}(2012)\citenamefont {Cuevas-Saavedra}, \citenamefont {Chakraborty}, \citenamefont {Rabi}, \citenamefont {C{\'a}rdenas},\ and\ \citenamefont {Ayers}}]{Saavedra12}%
  \BibitemOpen
  \bibfield  {author} {\bibinfo {author} {\bibfnamefont {R.}~\bibnamefont {Cuevas-Saavedra}}, \bibinfo {author} {\bibfnamefont {D.}~\bibnamefont {Chakraborty}}, \bibinfo {author} {\bibfnamefont {S.}~\bibnamefont {Rabi}}, \bibinfo {author} {\bibfnamefont {C.}~\bibnamefont {C{\'a}rdenas}},\ and\ \bibinfo {author} {\bibfnamefont {P.~W.}\ \bibnamefont {Ayers}},\ }\href@noop {} {\bibfield  {journal} {\bibinfo  {journal} {J. Chem. Theory Comput.}\ }\textbf {\bibinfo {volume} {8}},\ \bibinfo {pages} {4081} (\bibinfo {year} {2012})}\BibitemShut {NoStop}%
\bibitem [{\citenamefont {Perdew}\ \emph {et~al.}(1982)\citenamefont {Perdew}, \citenamefont {Parr}, \citenamefont {Levy},\ and\ \citenamefont {Balduz}}]{PPLB82}%
  \BibitemOpen
  \bibfield  {author} {\bibinfo {author} {\bibfnamefont {J.~P.}\ \bibnamefont {Perdew}}, \bibinfo {author} {\bibfnamefont {R.~G.}\ \bibnamefont {Parr}}, \bibinfo {author} {\bibfnamefont {M.}~\bibnamefont {Levy}},\ and\ \bibinfo {author} {\bibfnamefont {J.~L.}\ \bibnamefont {Balduz}},\ }\href@noop {} {\bibfield  {journal} {\bibinfo  {journal} {Phys. Rev. Lett.}\ }\textbf {\bibinfo {volume} {49}},\ \bibinfo {pages} {1691} (\bibinfo {year} {1982})}\BibitemShut {NoStop}%
\bibitem [{\citenamefont {Cohen}\ \emph {et~al.}(2008{\natexlab{a}})\citenamefont {Cohen}, \citenamefont {Mori-S\'{a}nchez},\ and\ \citenamefont {Yang}}]{CohenMoriSYang08}%
  \BibitemOpen
  \bibfield  {author} {\bibinfo {author} {\bibfnamefont {A.~J.}\ \bibnamefont {Cohen}}, \bibinfo {author} {\bibfnamefont {P.}~\bibnamefont {Mori-S\'{a}nchez}},\ and\ \bibinfo {author} {\bibfnamefont {W.}~\bibnamefont {Yang}},\ }\href@noop {} {\bibfield  {journal} {\bibinfo  {journal} {J. Chem. Phys.}\ }\textbf {\bibinfo {volume} {129}} (\bibinfo {year} {2008}{\natexlab{a}})}\BibitemShut {NoStop}%
\bibitem [{\citenamefont {Cohen}\ \emph {et~al.}(2008{\natexlab{b}})\citenamefont {Cohen}, \citenamefont {Mori-S\'{a}nchez},\ and\ \citenamefont {Yang}}]{Cohen08}%
  \BibitemOpen
  \bibfield  {author} {\bibinfo {author} {\bibfnamefont {A.~J.}\ \bibnamefont {Cohen}}, \bibinfo {author} {\bibfnamefont {P.}~\bibnamefont {Mori-S\'{a}nchez}},\ and\ \bibinfo {author} {\bibfnamefont {W.}~\bibnamefont {Yang}},\ }\href@noop {} {\bibfield  {journal} {\bibinfo  {journal} {Science}\ }\textbf {\bibinfo {volume} {321}},\ \bibinfo {pages} {792} (\bibinfo {year} {2008}{\natexlab{b}})}\BibitemShut {NoStop}%
\bibitem [{\citenamefont {Mori-S\'anchez}\ \emph {et~al.}(2009)\citenamefont {Mori-S\'anchez}, \citenamefont {Cohen},\ and\ \citenamefont {Yang}}]{MoriS09}%
  \BibitemOpen
  \bibfield  {author} {\bibinfo {author} {\bibfnamefont {P.}~\bibnamefont {Mori-S\'anchez}}, \bibinfo {author} {\bibfnamefont {A.~J.}\ \bibnamefont {Cohen}},\ and\ \bibinfo {author} {\bibfnamefont {W.}~\bibnamefont {Yang}},\ }\href {https://doi.org/10.1103/PhysRevLett.102.066403} {\bibfield  {journal} {\bibinfo  {journal} {Phys. Rev. Lett.}\ }\textbf {\bibinfo {volume} {102}},\ \bibinfo {pages} {066403} (\bibinfo {year} {2009})}\BibitemShut {NoStop}%
\bibitem [{\citenamefont {Cohen}\ \emph {et~al.}(2012)\citenamefont {Cohen}, \citenamefont {Mori-S\'{a}nchez},\ and\ \citenamefont {Yang}}]{cohen12}%
  \BibitemOpen
  \bibfield  {author} {\bibinfo {author} {\bibfnamefont {A.~J.}\ \bibnamefont {Cohen}}, \bibinfo {author} {\bibfnamefont {P.}~\bibnamefont {Mori-S\'{a}nchez}},\ and\ \bibinfo {author} {\bibfnamefont {W.}~\bibnamefont {Yang}},\ }\href@noop {} {\bibfield  {journal} {\bibinfo  {journal} {Chem. Rev.}\ }\textbf {\bibinfo {volume} {112}},\ \bibinfo {pages} {289} (\bibinfo {year} {2012})}\BibitemShut {NoStop}%
\bibitem [{\citenamefont {Su}\ \emph {et~al.}(2018)\citenamefont {Su}, \citenamefont {Li},\ and\ \citenamefont {Yang}}]{su2018describing}%
  \BibitemOpen
  \bibfield  {author} {\bibinfo {author} {\bibfnamefont {N.~Q.}\ \bibnamefont {Su}}, \bibinfo {author} {\bibfnamefont {C.}~\bibnamefont {Li}},\ and\ \bibinfo {author} {\bibfnamefont {W.}~\bibnamefont {Yang}},\ }\href@noop {} {\bibfield  {journal} {\bibinfo  {journal} {Proceedings of the National Academy of Sciences}\ }\textbf {\bibinfo {volume} {115}},\ \bibinfo {pages} {9678} (\bibinfo {year} {2018})}\BibitemShut {NoStop}%
\bibitem [{\citenamefont {Cohen}\ \emph {et~al.}(2009)\citenamefont {Cohen}, \citenamefont {Mori-Sanchez},\ and\ \citenamefont {Yang}}]{cohen_fractional_MP2_09}%
  \BibitemOpen
  \bibfield  {author} {\bibinfo {author} {\bibfnamefont {A.~J.}\ \bibnamefont {Cohen}}, \bibinfo {author} {\bibfnamefont {P.}~\bibnamefont {Mori-Sanchez}},\ and\ \bibinfo {author} {\bibfnamefont {W.}~\bibnamefont {Yang}},\ }\href@noop {} {\bibfield  {journal} {\bibinfo  {journal} {Journal of chemical theory and computation}\ }\textbf {\bibinfo {volume} {5}},\ \bibinfo {pages} {786} (\bibinfo {year} {2009})}\BibitemShut {NoStop}%
\bibitem [{\citenamefont {Burgess}\ \emph {et~al.}(2023)\citenamefont {Burgess}, \citenamefont {Linscott},\ and\ \citenamefont {O'Regan}}]{burgess2023dft+}%
  \BibitemOpen
  \bibfield  {author} {\bibinfo {author} {\bibfnamefont {A.~C.}\ \bibnamefont {Burgess}}, \bibinfo {author} {\bibfnamefont {E.}~\bibnamefont {Linscott}},\ and\ \bibinfo {author} {\bibfnamefont {D.~D.}\ \bibnamefont {O'Regan}},\ }\href@noop {} {\bibfield  {journal} {\bibinfo  {journal} {Physical Review B}\ }\textbf {\bibinfo {volume} {107}},\ \bibinfo {pages} {L121115} (\bibinfo {year} {2023})}\BibitemShut {NoStop}%
\bibitem [{\citenamefont {Bajaj}\ \emph {et~al.}(2017)\citenamefont {Bajaj}, \citenamefont {Janet},\ and\ \citenamefont {Kulik}}]{BajajKulik17}%
  \BibitemOpen
  \bibfield  {author} {\bibinfo {author} {\bibfnamefont {A.}~\bibnamefont {Bajaj}}, \bibinfo {author} {\bibfnamefont {J.~P.}\ \bibnamefont {Janet}},\ and\ \bibinfo {author} {\bibfnamefont {H.~J.}\ \bibnamefont {Kulik}},\ }\href@noop {} {\bibfield  {journal} {\bibinfo  {journal} {J. Chem. Phys.}\ }\textbf {\bibinfo {volume} {147}},\ \bibinfo {pages} {191101} (\bibinfo {year} {2017})}\BibitemShut {NoStop}%
\bibitem [{\citenamefont {Bajaj}\ \emph {et~al.}(2022)\citenamefont {Bajaj}, \citenamefont {Duan}, \citenamefont {Nandy}, \citenamefont {Taylor},\ and\ \citenamefont {Kulik}}]{BajajKulik22}%
  \BibitemOpen
  \bibfield  {author} {\bibinfo {author} {\bibfnamefont {A.}~\bibnamefont {Bajaj}}, \bibinfo {author} {\bibfnamefont {C.}~\bibnamefont {Duan}}, \bibinfo {author} {\bibfnamefont {A.}~\bibnamefont {Nandy}}, \bibinfo {author} {\bibfnamefont {M.~G.}\ \bibnamefont {Taylor}},\ and\ \bibinfo {author} {\bibfnamefont {H.~J.}\ \bibnamefont {Kulik}},\ }\href@noop {} {\bibfield  {journal} {\bibinfo  {journal} {J. Chem. Phys.}\ }\textbf {\bibinfo {volume} {156}},\ \bibinfo {pages} {184112} (\bibinfo {year} {2022})}\BibitemShut {NoStop}%
\bibitem [{\citenamefont {De~Vriendt}\ \emph {et~al.}(2021)\citenamefont {De~Vriendt}, \citenamefont {Lemmens}, \citenamefont {De~Baerdemacker}, \citenamefont {Bultinck},\ and\ \citenamefont {Acke}}]{de2021quantifying}%
  \BibitemOpen
  \bibfield  {author} {\bibinfo {author} {\bibfnamefont {X.}~\bibnamefont {De~Vriendt}}, \bibinfo {author} {\bibfnamefont {L.}~\bibnamefont {Lemmens}}, \bibinfo {author} {\bibfnamefont {S.}~\bibnamefont {De~Baerdemacker}}, \bibinfo {author} {\bibfnamefont {P.}~\bibnamefont {Bultinck}},\ and\ \bibinfo {author} {\bibfnamefont {G.}~\bibnamefont {Acke}},\ }\href@noop {} {\bibfield  {journal} {\bibinfo  {journal} {Journal of Chemical Theory and Computation}\ }\textbf {\bibinfo {volume} {17}},\ \bibinfo {pages} {6808} (\bibinfo {year} {2021})}\BibitemShut {NoStop}%
\bibitem [{\citenamefont {Prokopiou}\ \emph {et~al.}(2022)\citenamefont {Prokopiou}, \citenamefont {Hartstein}, \citenamefont {Govind},\ and\ \citenamefont {Kronik}}]{Prokopiu22}%
  \BibitemOpen
  \bibfield  {author} {\bibinfo {author} {\bibfnamefont {G.}~\bibnamefont {Prokopiou}}, \bibinfo {author} {\bibfnamefont {M.}~\bibnamefont {Hartstein}}, \bibinfo {author} {\bibfnamefont {N.}~\bibnamefont {Govind}},\ and\ \bibinfo {author} {\bibfnamefont {L.}~\bibnamefont {Kronik}},\ }\href@noop {} {\bibfield  {journal} {\bibinfo  {journal} {J. Chem. Theory Comput.}\ }\textbf {\bibinfo {volume} {18}},\ \bibinfo {pages} {2331} (\bibinfo {year} {2022})}\BibitemShut {NoStop}%
\bibitem [{\citenamefont {Wodyński}\ \emph {et~al.}(2021)\citenamefont {Wodyński}, \citenamefont {Arbuznikov},\ and\ \citenamefont {Kaupp}}]{WodynskiArbuzKaupp21}%
  \BibitemOpen
  \bibfield  {author} {\bibinfo {author} {\bibfnamefont {A.}~\bibnamefont {Wodyński}}, \bibinfo {author} {\bibfnamefont {A.~V.}\ \bibnamefont {Arbuznikov}},\ and\ \bibinfo {author} {\bibfnamefont {M.}~\bibnamefont {Kaupp}},\ }\href@noop {} {\bibfield  {journal} {\bibinfo  {journal} {J. Chem. Phys.}\ }\textbf {\bibinfo {volume} {155}},\ \bibinfo {pages} {144101} (\bibinfo {year} {2021})}\BibitemShut {NoStop}%
\bibitem [{\citenamefont {Yang}\ \emph {et~al.}(2016)\citenamefont {Yang}, \citenamefont {Patel}, \citenamefont {Miranda-Quintana}, \citenamefont {Heidar-Zadeh}, \citenamefont {Gonz\'{a}lez-Espinoza},\ and\ \citenamefont {Ayers}}]{XDYang16}%
  \BibitemOpen
  \bibfield  {author} {\bibinfo {author} {\bibfnamefont {X.~D.}\ \bibnamefont {Yang}}, \bibinfo {author} {\bibfnamefont {A.~H.~G.}\ \bibnamefont {Patel}}, \bibinfo {author} {\bibfnamefont {R.~A.}\ \bibnamefont {Miranda-Quintana}}, \bibinfo {author} {\bibfnamefont {F.}~\bibnamefont {Heidar-Zadeh}}, \bibinfo {author} {\bibfnamefont {C.~E.}\ \bibnamefont {Gonz\'{a}lez-Espinoza}},\ and\ \bibinfo {author} {\bibfnamefont {P.~W.}\ \bibnamefont {Ayers}},\ }\href@noop {} {\bibfield  {journal} {\bibinfo  {journal} {J. Chem. Phys.}\ }\textbf {\bibinfo {volume} {145}} (\bibinfo {year} {2016})}\BibitemShut {NoStop}%
\bibitem [{\citenamefont {Yang}\ \emph {et~al.}(2000)\citenamefont {Yang}, \citenamefont {Zhang},\ and\ \citenamefont {Ayers}}]{Yang00}%
  \BibitemOpen
  \bibfield  {author} {\bibinfo {author} {\bibfnamefont {W.}~\bibnamefont {Yang}}, \bibinfo {author} {\bibfnamefont {Y.}~\bibnamefont {Zhang}},\ and\ \bibinfo {author} {\bibfnamefont {P.~W.}\ \bibnamefont {Ayers}},\ }\href@noop {} {\bibfield  {journal} {\bibinfo  {journal} {Phys. Rev. Lett.}\ }\textbf {\bibinfo {volume} {84}},\ \bibinfo {pages} {5172} (\bibinfo {year} {2000})}\BibitemShut {NoStop}%
\bibitem [{\citenamefont {Chan}(1999)}]{Chan99}%
  \BibitemOpen
  \bibfield  {author} {\bibinfo {author} {\bibfnamefont {G.~K.-L.}\ \bibnamefont {Chan}},\ }\href@noop {} {\bibfield  {journal} {\bibinfo  {journal} {J. Chem. Phys.}\ }\textbf {\bibinfo {volume} {110}},\ \bibinfo {pages} {4710} (\bibinfo {year} {1999})}\BibitemShut {NoStop}%
\bibitem [{\citenamefont {Goshen}\ and\ \citenamefont {Kraisler}(2024)}]{GoshenKraisler24}%
  \BibitemOpen
  \bibfield  {author} {\bibinfo {author} {\bibfnamefont {Y.}~\bibnamefont {Goshen}}\ and\ \bibinfo {author} {\bibfnamefont {E.}~\bibnamefont {Kraisler}},\ }\href@noop {} {\bibfield  {journal} {\bibinfo  {journal} {J. Phys. Chem. Lett.}\ }\textbf {\bibinfo {volume} {15}},\ \bibinfo {pages} {2337} (\bibinfo {year} {2024})}\BibitemShut {NoStop}%
\bibitem [{\citenamefont {Burgess}\ \emph {et~al.}(2024)\citenamefont {Burgess}, \citenamefont {Linscott},\ and\ \citenamefont {O'Regan}}]{BurgessLinscottORegan24}%
  \BibitemOpen
  \bibfield  {author} {\bibinfo {author} {\bibfnamefont {A.~C.}\ \bibnamefont {Burgess}}, \bibinfo {author} {\bibfnamefont {E.}~\bibnamefont {Linscott}},\ and\ \bibinfo {author} {\bibfnamefont {D.~D.}\ \bibnamefont {O'Regan}},\ }\href@noop {} {\bibfield  {journal} {\bibinfo  {journal} {Phys. Rev. Lett.}\ }\textbf {\bibinfo {volume} {133}},\ \bibinfo {pages} {026404} (\bibinfo {year} {2024})}\BibitemShut {NoStop}%
\bibitem [{\citenamefont {Capelle}\ \emph {et~al.}(2010)\citenamefont {Capelle}, \citenamefont {Vignale},\ and\ \citenamefont {Ullrich}}]{CapVigUll10}%
  \BibitemOpen
  \bibfield  {author} {\bibinfo {author} {\bibfnamefont {K.}~\bibnamefont {Capelle}}, \bibinfo {author} {\bibfnamefont {G.}~\bibnamefont {Vignale}},\ and\ \bibinfo {author} {\bibfnamefont {C.~A.}\ \bibnamefont {Ullrich}},\ }\href@noop {} {\bibfield  {journal} {\bibinfo  {journal} {Phys. Rev. B}\ }\textbf {\bibinfo {volume} {81}},\ \bibinfo {pages} {125114} (\bibinfo {year} {2010})}\BibitemShut {NoStop}%
\bibitem [{\citenamefont {Gould}\ and\ \citenamefont {Dobson}(2013)}]{GouldDobson13}%
  \BibitemOpen
  \bibfield  {author} {\bibinfo {author} {\bibfnamefont {T.}~\bibnamefont {Gould}}\ and\ \bibinfo {author} {\bibfnamefont {J.~F.}\ \bibnamefont {Dobson}},\ }\href@noop {} {\bibfield  {journal} {\bibinfo  {journal} {J. Chem. Phys.}\ }\textbf {\bibinfo {volume} {138}},\ \bibinfo {pages} {014103} (\bibinfo {year} {2013})}\BibitemShut {NoStop}%
\bibitem [{\citenamefont {Kronik}\ \emph {et~al.}(2012)\citenamefont {Kronik}, \citenamefont {Stein}, \citenamefont {Refaely-Abramson},\ and\ \citenamefont {Baer}}]{Kronik_JCTC_review12}%
  \BibitemOpen
  \bibfield  {author} {\bibinfo {author} {\bibfnamefont {L.}~\bibnamefont {Kronik}}, \bibinfo {author} {\bibfnamefont {T.}~\bibnamefont {Stein}}, \bibinfo {author} {\bibfnamefont {S.}~\bibnamefont {Refaely-Abramson}},\ and\ \bibinfo {author} {\bibfnamefont {R.}~\bibnamefont {Baer}},\ }\href@noop {} {\bibfield  {journal} {\bibinfo  {journal} {J. Chem. Theory Comput.}\ }\textbf {\bibinfo {volume} {8}},\ \bibinfo {pages} {1515} (\bibinfo {year} {2012})}\BibitemShut {NoStop}%
\bibitem [{\citenamefont {Stein}\ \emph {et~al.}(2012)\citenamefont {Stein}, \citenamefont {Autschbach}, \citenamefont {Govind}, \citenamefont {Kronik},\ and\ \citenamefont {Baer}}]{SteinKronikBaer_curvatures12}%
  \BibitemOpen
  \bibfield  {author} {\bibinfo {author} {\bibfnamefont {T.}~\bibnamefont {Stein}}, \bibinfo {author} {\bibfnamefont {J.}~\bibnamefont {Autschbach}}, \bibinfo {author} {\bibfnamefont {N.}~\bibnamefont {Govind}}, \bibinfo {author} {\bibfnamefont {L.}~\bibnamefont {Kronik}},\ and\ \bibinfo {author} {\bibfnamefont {R.}~\bibnamefont {Baer}},\ }\href@noop {} {\bibfield  {journal} {\bibinfo  {journal} {J. Phys. Chem. Lett.}\ }\textbf {\bibinfo {volume} {3}},\ \bibinfo {pages} {3740} (\bibinfo {year} {2012})}\BibitemShut {NoStop}%
\bibitem [{\citenamefont {Stein}\ \emph {et~al.}(2009)\citenamefont {Stein}, \citenamefont {Kronik},\ and\ \citenamefont {Baer}}]{stein2009reliable}%
  \BibitemOpen
  \bibfield  {author} {\bibinfo {author} {\bibfnamefont {T.}~\bibnamefont {Stein}}, \bibinfo {author} {\bibfnamefont {L.}~\bibnamefont {Kronik}},\ and\ \bibinfo {author} {\bibfnamefont {R.}~\bibnamefont {Baer}},\ }\href@noop {} {\bibfield  {journal} {\bibinfo  {journal} {Journal of the American Chemical Society}\ }\textbf {\bibinfo {volume} {131}},\ \bibinfo {pages} {2818} (\bibinfo {year} {2009})}\BibitemShut {NoStop}%
\bibitem [{\citenamefont {Stein}\ \emph {et~al.}(2010)\citenamefont {Stein}, \citenamefont {Eisenberg}, \citenamefont {Kronik},\ and\ \citenamefont {Baer}}]{Stein10}%
  \BibitemOpen
  \bibfield  {author} {\bibinfo {author} {\bibfnamefont {T.}~\bibnamefont {Stein}}, \bibinfo {author} {\bibfnamefont {H.}~\bibnamefont {Eisenberg}}, \bibinfo {author} {\bibfnamefont {L.}~\bibnamefont {Kronik}},\ and\ \bibinfo {author} {\bibfnamefont {R.}~\bibnamefont {Baer}},\ }\href@noop {} {\bibfield  {journal} {\bibinfo  {journal} {Phys. Rev. Lett.}\ }\textbf {\bibinfo {volume} {105}},\ \bibinfo {pages} {266802} (\bibinfo {year} {2010})}\BibitemShut {NoStop}%
\bibitem [{\citenamefont {Kraisler}\ and\ \citenamefont {Kronik}(2013)}]{KraislerKronik13}%
  \BibitemOpen
  \bibfield  {author} {\bibinfo {author} {\bibfnamefont {E.}~\bibnamefont {Kraisler}}\ and\ \bibinfo {author} {\bibfnamefont {L.}~\bibnamefont {Kronik}},\ }\href@noop {} {\bibfield  {journal} {\bibinfo  {journal} {Phys. Rev. Lett.}\ }\textbf {\bibinfo {volume} {110}},\ \bibinfo {pages} {126403} (\bibinfo {year} {2013})}\BibitemShut {NoStop}%
\bibitem [{\citenamefont {Kraisler}\ and\ \citenamefont {Kronik}(2014)}]{KraislerKronik14}%
  \BibitemOpen
  \bibfield  {author} {\bibinfo {author} {\bibfnamefont {E.}~\bibnamefont {Kraisler}}\ and\ \bibinfo {author} {\bibfnamefont {L.}~\bibnamefont {Kronik}},\ }\href@noop {} {\bibfield  {journal} {\bibinfo  {journal} {J. Chem. Phys.}\ }\textbf {\bibinfo {volume} {140}},\ \bibinfo {pages} {18A540} (\bibinfo {year} {2014})}\BibitemShut {NoStop}%
\bibitem [{\citenamefont {Kraisler}\ \emph {et~al.}(2015)\citenamefont {Kraisler}, \citenamefont {Schmidt}, \citenamefont {K\"{u}mmel},\ and\ \citenamefont {Kronik}}]{KraislerSchmidt15}%
  \BibitemOpen
  \bibfield  {author} {\bibinfo {author} {\bibfnamefont {E.}~\bibnamefont {Kraisler}}, \bibinfo {author} {\bibfnamefont {T.}~\bibnamefont {Schmidt}}, \bibinfo {author} {\bibfnamefont {S.}~\bibnamefont {K\"{u}mmel}},\ and\ \bibinfo {author} {\bibfnamefont {L.}~\bibnamefont {Kronik}},\ }\href@noop {} {\bibfield  {journal} {\bibinfo  {journal} {J. Chem. Phys.}\ }\textbf {\bibinfo {volume} {143}},\ \bibinfo {pages} {104105} (\bibinfo {year} {2015})}\BibitemShut {NoStop}%
\bibitem [{\citenamefont {Kraisler}\ and\ \citenamefont {Kronik}(2015)}]{KraislerKronik15}%
  \BibitemOpen
  \bibfield  {author} {\bibinfo {author} {\bibfnamefont {E.}~\bibnamefont {Kraisler}}\ and\ \bibinfo {author} {\bibfnamefont {L.}~\bibnamefont {Kronik}},\ }\href@noop {} {\bibfield  {journal} {\bibinfo  {journal} {Phys. Rev. A}\ }\textbf {\bibinfo {volume} {91}},\ \bibinfo {pages} {032504} (\bibinfo {year} {2015})}\BibitemShut {NoStop}%
\bibitem [{NIS()}]{NIST_lines}%
  \BibitemOpen
  \href@noop {} {}\bibinfo {note} {{https://physics.nist.gov/PhysRefData/ASD/levels\_form.html}}\BibitemShut {NoStop}%
\bibitem [{\citenamefont {Janak}(1978)}]{Janak78}%
  \BibitemOpen
  \bibfield  {author} {\bibinfo {author} {\bibfnamefont {J.~F.}\ \bibnamefont {Janak}},\ }\href@noop {} {\bibfield  {journal} {\bibinfo  {journal} {Phys. Rev. B}\ }\textbf {\bibinfo {volume} {18}},\ \bibinfo {pages} {7165} (\bibinfo {year} {1978})}\BibitemShut {NoStop}%
\bibitem [{Note1()}]{Note1}%
  \BibitemOpen
  \bibinfo {note} {In atomic calculations within the spherical approximation, it is common to plot not the density itself, but the quantity $4 \pi r^2 n(\protect \mathbf {r})$. This quantity is easier to visualize and, among other advantages, reveals the shell structure of the density. The same applies also to $ d^\sigma (\protect \mathbf {r};x)$ in the present context.}\BibitemShut {Stop}%
\bibitem [{\citenamefont {Kraisler}\ \emph {et~al.}(2009)\citenamefont {Kraisler}, \citenamefont {Makov}, \citenamefont {Argaman},\ and\ \citenamefont {Kelson}}]{KraislerMakovArgamanKelson09}%
  \BibitemOpen
  \bibfield  {author} {\bibinfo {author} {\bibfnamefont {E.}~\bibnamefont {Kraisler}}, \bibinfo {author} {\bibfnamefont {G.}~\bibnamefont {Makov}}, \bibinfo {author} {\bibfnamefont {N.}~\bibnamefont {Argaman}},\ and\ \bibinfo {author} {\bibfnamefont {I.}~\bibnamefont {Kelson}},\ }\href@noop {} {\bibfield  {journal} {\bibinfo  {journal} {Phys. Rev. A}\ }\textbf {\bibinfo {volume} {80}},\ \bibinfo {pages} {032115} (\bibinfo {year} {2009})}\BibitemShut {NoStop}%
\bibitem [{\citenamefont {Kraisler}\ \emph {et~al.}(2010)\citenamefont {Kraisler}, \citenamefont {Makov},\ and\ \citenamefont {Kelson}}]{KraislerMakovKelson10}%
  \BibitemOpen
  \bibfield  {author} {\bibinfo {author} {\bibfnamefont {E.}~\bibnamefont {Kraisler}}, \bibinfo {author} {\bibfnamefont {G.}~\bibnamefont {Makov}},\ and\ \bibinfo {author} {\bibfnamefont {I.}~\bibnamefont {Kelson}},\ }\href@noop {} {\bibfield  {journal} {\bibinfo  {journal} {Phys. Rev. A}\ }\textbf {\bibinfo {volume} {82}},\ \bibinfo {pages} {042516} (\bibinfo {year} {2010})}\BibitemShut {NoStop}%
\bibitem [{\citenamefont {Krieger}\ \emph {et~al.}(1992)\citenamefont {Krieger}, \citenamefont {Li},\ and\ \citenamefont {Iafrate}}]{KLI92}%
  \BibitemOpen
  \bibfield  {author} {\bibinfo {author} {\bibfnamefont {J.~B.}\ \bibnamefont {Krieger}}, \bibinfo {author} {\bibfnamefont {Y.}~\bibnamefont {Li}},\ and\ \bibinfo {author} {\bibfnamefont {G.~J.}\ \bibnamefont {Iafrate}},\ }\href@noop {} {\bibfield  {journal} {\bibinfo  {journal} {Phys. Rev. A}\ }\textbf {\bibinfo {volume} {45}},\ \bibinfo {pages} {101} (\bibinfo {year} {1992})}\BibitemShut {NoStop}%
\bibitem [{\citenamefont {K\"ummel}\ and\ \citenamefont {Kronik}(2008)}]{KK08}%
  \BibitemOpen
  \bibfield  {author} {\bibinfo {author} {\bibfnamefont {S.}~\bibnamefont {K\"ummel}}\ and\ \bibinfo {author} {\bibfnamefont {L.}~\bibnamefont {Kronik}},\ }\href@noop {} {\bibfield  {journal} {\bibinfo  {journal} {Rev. Mod. Phys.}\ }\textbf {\bibinfo {volume} {80}},\ \bibinfo {pages} {3} (\bibinfo {year} {2008})}\BibitemShut {NoStop}%
\bibitem [{\citenamefont {Makmal}\ \emph {et~al.}(2009)\citenamefont {Makmal}, \citenamefont {K\"ummel},\ and\ \citenamefont {Kronik}}]{Makmal09JCTC}%
  \BibitemOpen
  \bibfield  {author} {\bibinfo {author} {\bibfnamefont {A.}~\bibnamefont {Makmal}}, \bibinfo {author} {\bibfnamefont {S.}~\bibnamefont {K\"ummel}},\ and\ \bibinfo {author} {\bibfnamefont {L.}~\bibnamefont {Kronik}},\ }\href@noop {} {\bibfield  {journal} {\bibinfo  {journal} {J. Chem. Theory Comput.}\ }\textbf {\bibinfo {volume} {5}},\ \bibinfo {pages} {1731} (\bibinfo {year} {2009})}\BibitemShut {NoStop}%
\bibitem [{\citenamefont {Hodgson}\ \emph {et~al.}(2017)\citenamefont {Hodgson}, \citenamefont {Kraisler}, \citenamefont {Schild},\ and\ \citenamefont {Gross}}]{HodgsonKraisler17}%
  \BibitemOpen
  \bibfield  {author} {\bibinfo {author} {\bibfnamefont {M.~J.~P.}\ \bibnamefont {Hodgson}}, \bibinfo {author} {\bibfnamefont {E.}~\bibnamefont {Kraisler}}, \bibinfo {author} {\bibfnamefont {A.}~\bibnamefont {Schild}},\ and\ \bibinfo {author} {\bibfnamefont {E.~K.~U.}\ \bibnamefont {Gross}},\ }\href@noop {} {\bibfield  {journal} {\bibinfo  {journal} {J. Phys. Chem. Lett.}\ }\textbf {\bibinfo {volume} {8}},\ \bibinfo {pages} {5974} (\bibinfo {year} {2017})}\BibitemShut {NoStop}%
\bibitem [{\citenamefont {Kraisler}\ \emph {et~al.}(2021)\citenamefont {Kraisler}, \citenamefont {Hodgson},\ and\ \citenamefont {Gross}}]{Kraisler21}%
  \BibitemOpen
  \bibfield  {author} {\bibinfo {author} {\bibfnamefont {E.}~\bibnamefont {Kraisler}}, \bibinfo {author} {\bibfnamefont {M.~J.~P.}\ \bibnamefont {Hodgson}},\ and\ \bibinfo {author} {\bibfnamefont {E.~K.~U.}\ \bibnamefont {Gross}},\ }\href@noop {} {\bibfield  {journal} {\bibinfo  {journal} {J. Chem. Theory Comput.}\ }\textbf {\bibinfo {volume} {17}},\ \bibinfo {pages} {1390} (\bibinfo {year} {2021})}\BibitemShut {NoStop}%
\bibitem [{\citenamefont {Kim}\ \emph {et~al.}(2013)\citenamefont {Kim}, \citenamefont {Sim},\ and\ \citenamefont {Burke}}]{kim2013understanding}%
  \BibitemOpen
  \bibfield  {author} {\bibinfo {author} {\bibfnamefont {M.-C.}\ \bibnamefont {Kim}}, \bibinfo {author} {\bibfnamefont {E.}~\bibnamefont {Sim}},\ and\ \bibinfo {author} {\bibfnamefont {K.}~\bibnamefont {Burke}},\ }\href@noop {} {\bibfield  {journal} {\bibinfo  {journal} {Phys. Rev. Lett.}\ }\textbf {\bibinfo {volume} {111}},\ \bibinfo {pages} {073003} (\bibinfo {year} {2013})}\BibitemShut {NoStop}%
\end{thebibliography}%


\begin{thebibliography}{0}%
\makeatletter
\providecommand \@ifxundefined [1]{%
 \@ifx{#1\undefined}
}%
\providecommand \@ifnum [1]{%
 \ifnum #1\expandafter \@firstoftwo
 \else \expandafter \@secondoftwo
 \fi
}%
\providecommand \@ifx [1]{%
 \ifx #1\expandafter \@firstoftwo
 \else \expandafter \@secondoftwo
 \fi
}%
\providecommand \natexlab [1]{#1}%
\providecommand \enquote  [1]{``#1''}%
\providecommand \bibnamefont  [1]{#1}%
\providecommand \bibfnamefont [1]{#1}%
\providecommand \citenamefont [1]{#1}%
\providecommand \href@noop [0]{\@secondoftwo}%
\providecommand \href [0]{\begingroup \@sanitize@url \@href}%
\providecommand \@href[1]{\@@startlink{#1}\@@href}%
\providecommand \@@href[1]{\endgroup#1\@@endlink}%
\providecommand \@sanitize@url [0]{\catcode `\\12\catcode `\$12\catcode `\&12\catcode `\#12\catcode `\^12\catcode `\_12\catcode `\%12\relax}%
\providecommand \@@startlink[1]{}%
\providecommand \@@endlink[0]{}%
\providecommand \url  [0]{\begingroup\@sanitize@url \@url }%
\providecommand \@url [1]{\endgroup\@href {#1}{\urlprefix }}%
\providecommand \urlprefix  [0]{URL }%
\providecommand \Eprint [0]{\href }%
\providecommand \doibase [0]{https://doi.org/}%
\providecommand \selectlanguage [0]{\@gobble}%
\providecommand \bibinfo  [0]{\@secondoftwo}%
\providecommand \bibfield  [0]{\@secondoftwo}%
\providecommand \translation [1]{[#1]}%
\providecommand \BibitemOpen [0]{}%
\providecommand \bibitemStop [0]{}%
\providecommand \bibitemNoStop [0]{.\EOS\space}%
\providecommand \EOS [0]{\spacefactor3000\relax}%
\providecommand \BibitemShut  [1]{\csname bibitem#1\endcsname}%
\let\auto@bib@innerbib\@empty
\end{thebibliography}%

\end{document}


\title{Supplementary Material: Spin migration in density functional theory: energy, potential and density perspectives}

\author{Alon Hayman}
\affiliation{Fritz Haber Research Center for Molecular Dynamics and Institute of Chemistry, The Hebrew University of Jerusalem, 9091401 Jerusalem, Israel}

\author{Eli Kraisler}
\email[Author to whom correspondence should be addressed: ]{eli.kraisler@mail.huji.ac.il}
\affiliation{Fritz Haber Research Center for Molecular Dynamics and Institute of Chemistry, The Hebrew University of Jerusalem, 9091401 Jerusalem, Israel}

\author{Tamar Stein}
\email[Author to whom correspondence should be addressed: ]{tamar.stein@mail.huji.ac.il}
\affiliation{Fritz Haber Research Center for Molecular Dynamics and Institute of Chemistry, The Hebrew University of Jerusalem, 9091401 Jerusalem, Israel}
\date{\today}

\maketitle{}

\mbox{}




This document provides the following supplementary material to the data presented in the main text. Additional examples are shown for all cases discussed in the main text in the different sections: Sec.~\ref{SI:sec:CaseI} here corresponds to Sec.~IV A of the main text: all-concave energy behavior, Sec.~\ref{SI:sec:CaseII} here corresponds to Sec.~IV B of the main text: mixed energy behavior, and likewise, Sec.~\ref{SI:sec:CaseIII} here corresponds to Sec.~IV C of the main text: all-convex energy behavior. 

In order to easily compare deviations in different approximate functionals, we present the total energy deviations, $\Delta E(S)$, which have been obtained from the raw $E(S)$ data by subtracting a straight line (see full explanation in the main text).

\subsection{CASE I} \label{SI:sec:CaseI}

Figure~1 of the main text presents $\Delta E(S)$ for the three systems discussed in Case I (H, Li and Na). In Table~\ref{stab:Case_I_energy_comp}, values of $\Delta E(S)$ at the peak (where $S=0$) of the aforementioned graphs are shown and analyzed in detail. Taking the linear combination of the EXX and PBE deviations, with weights of $\tfrac{1}{4}$ and $\tfrac{3}{4}$, respectively, provides a close approximation to the PBE0 value. 

\begin{table}[ht]
    \centering
    \begin{tabular}{@{\extracolsep{4pt}} l *{3}{c}@{}}
        \toprule
        \diagbox{Functional}{Atom} & \textbf{H} & \textbf{Li} & \textbf{Na} \\ 
        \midrule
        LSDA & 0.8992 & 0.2360 & 0.2026 \\ 
        PBE & 1.1173 & 0.2941 & 0.2202 \\ 
        EXX & 3.8719 & 1.4882 & 1.3797 \\ 
        PBE0 & 1.7744 & 0.5719 & 0.5044 \\ 
        $\tfrac{1}{4} \textrm{EXX} + \tfrac{3}{4} \textrm{PBE}$ &  1.8060 & 0.5926  & 0.5101 \\
        Relative difference vs.\ PBE0                           &  2\%       &    4\%  &   1\% \\
        \bottomrule
    \end{tabular}
    \caption{Deviation of the total energy, $\Delta E(S)$, at $S=0$ for H, Li and Na, for the functionals LSDA, PBE, EXX, and PBE0. The last two lines compare a linear combination of EXX and PBE to PBE0.} 
    \label{stab:Case_I_energy_comp} 
\end{table}

\begin{figure}[h]
    \centering
   \includegraphics[width=0.4\linewidth]{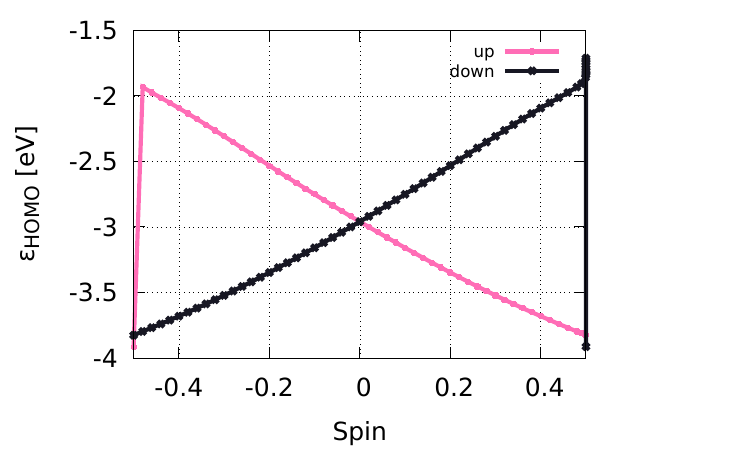}
    \caption{Frontier orbital energies for the Li atom versus the fractional spin using the LSDA0 functional.}
    \label{sfig:Li_HOMO_LDA0}
\end{figure}

\begin{figure*}
    \centering
    \hspace{-10mm}
\includegraphics[width=0.29\linewidth]{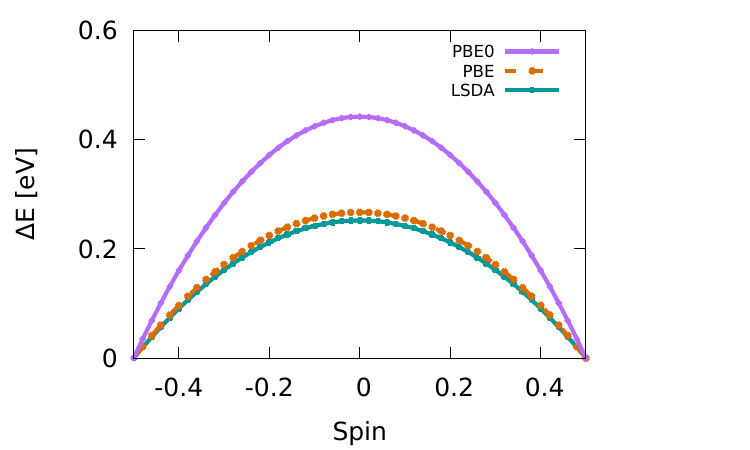}
    \hspace{-10mm}
\includegraphics[width=0.29\linewidth]{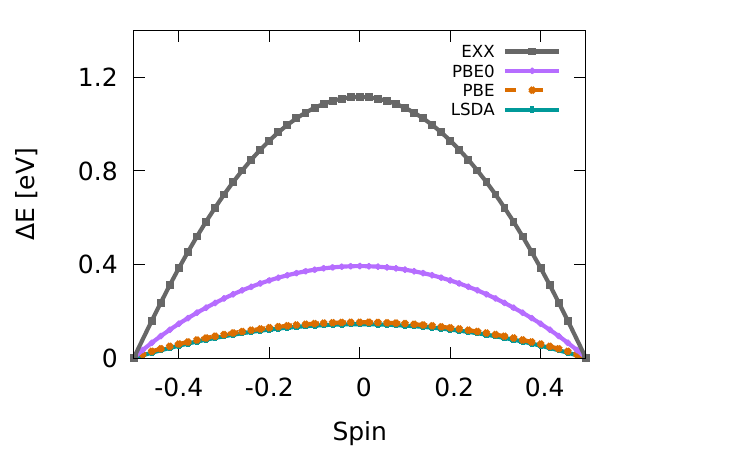}\\
    \hspace{-10mm}
\includegraphics[width=0.29\linewidth]{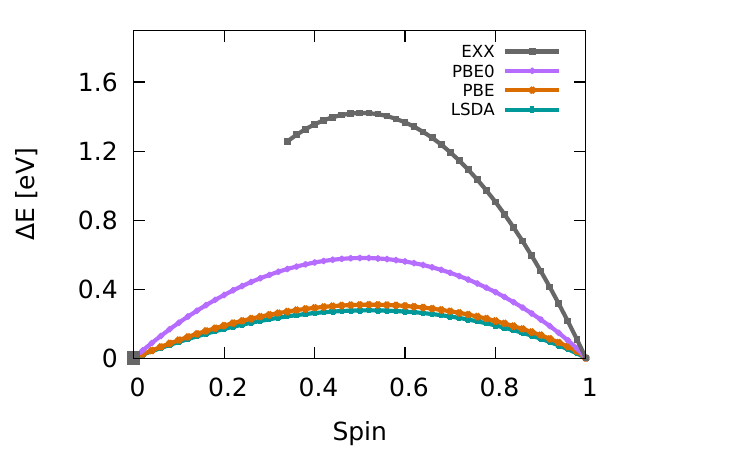}
    \hspace{-10mm}
\includegraphics[width=0.29\linewidth]{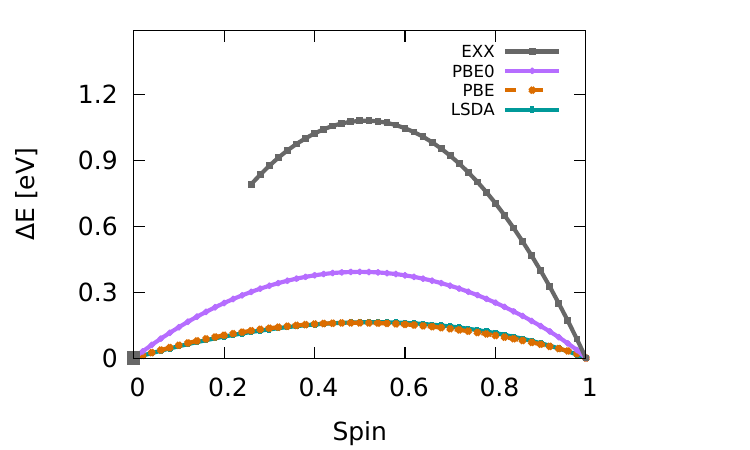}
    \hspace{-10mm}
\includegraphics[width=0.29\linewidth]{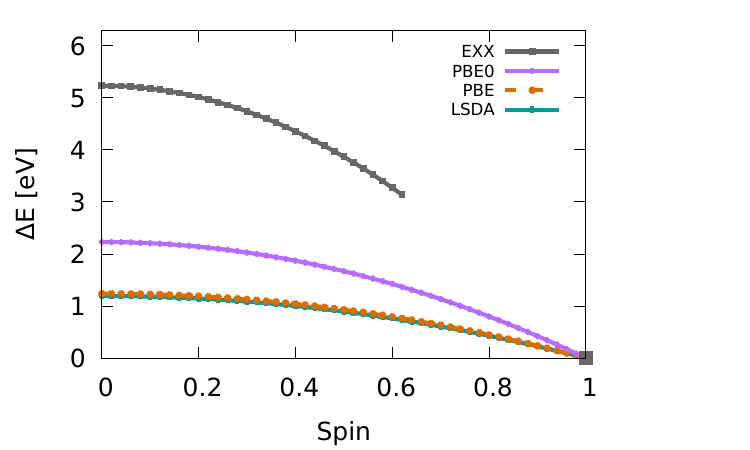}
    \hspace{-10mm}
\includegraphics[width=0.29\linewidth]{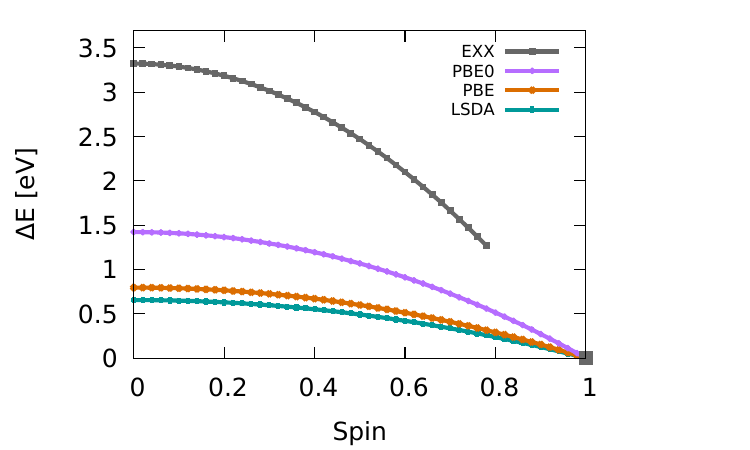}
    \hspace{-10mm}
    \caption{Deviation in energy, $\Delta E$, versus the fractional spin, $S$, for the B, K (top: left to right), Be, Mg, C, and Si atoms (bottom: left to right), for different XC approximations (see Legend).}
    \label{sfig:Be_Mg_C_SiTotal_E}
\end{figure*}

\begin{figure*}
    \centering
    \hspace{-12mm}
    \includegraphics[width=0.30\linewidth]{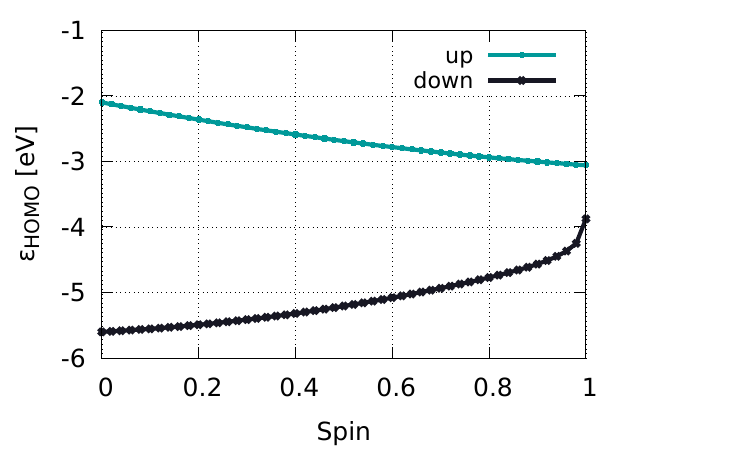}
    \hspace{-12mm}
   \includegraphics[width=0.30\linewidth]{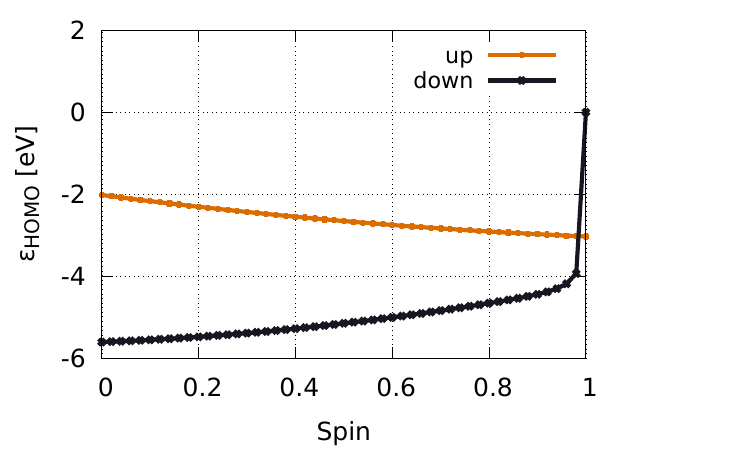}
    \hspace{-12mm}
   \includegraphics[width=0.30\linewidth]{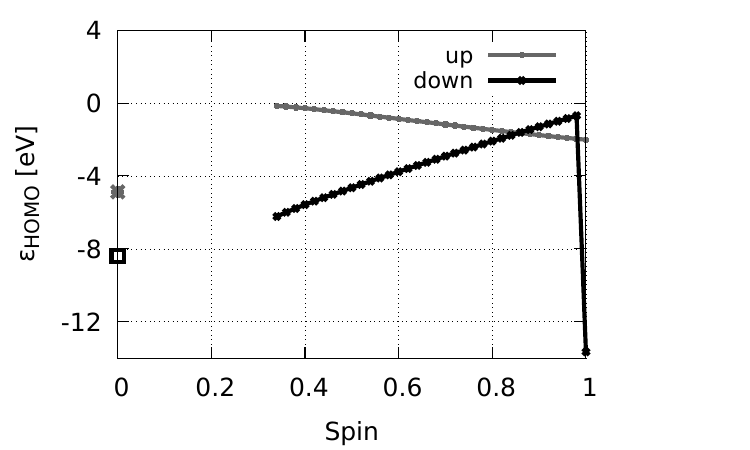}
   \hspace{-12mm}
   \includegraphics[width=0.30\linewidth]{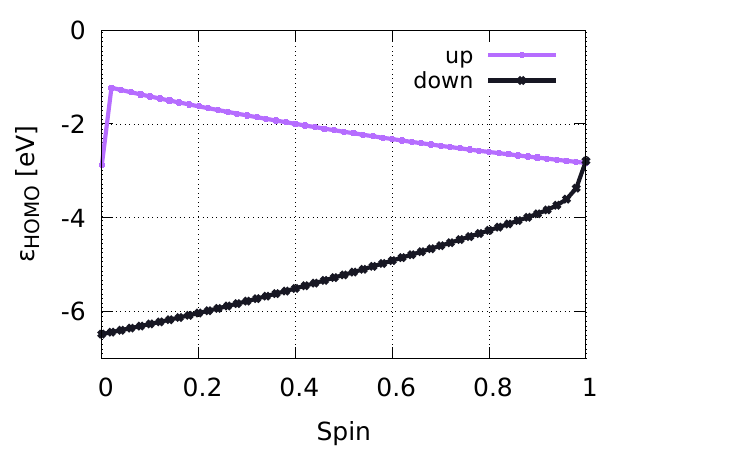}
    \hspace{-12mm}
    \caption{Frontier orbital energies for the Be atom versus the fractional spin using different XC functionals. From left to right: LSDA, PBE, EXX, PBE0.}
    \label{sfig:Be_HOMO}
\end{figure*}

\begin{figure*}
    \centering
    \hspace{-12mm}
    \includegraphics[width=0.30\linewidth]{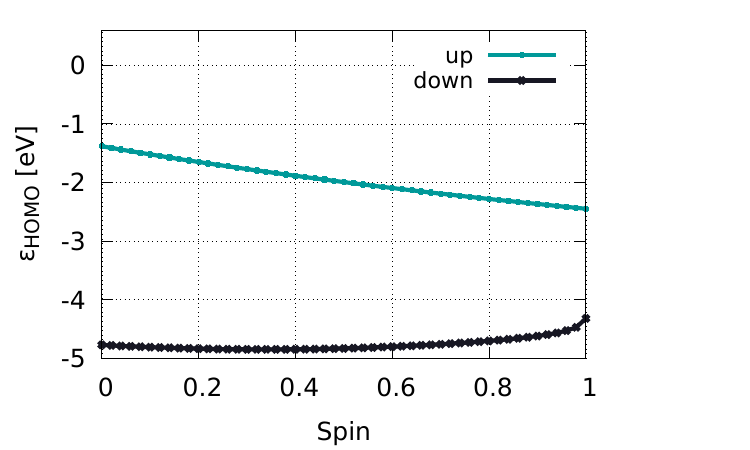}
    \hspace{-12mm}
   \includegraphics[width=0.30\linewidth]{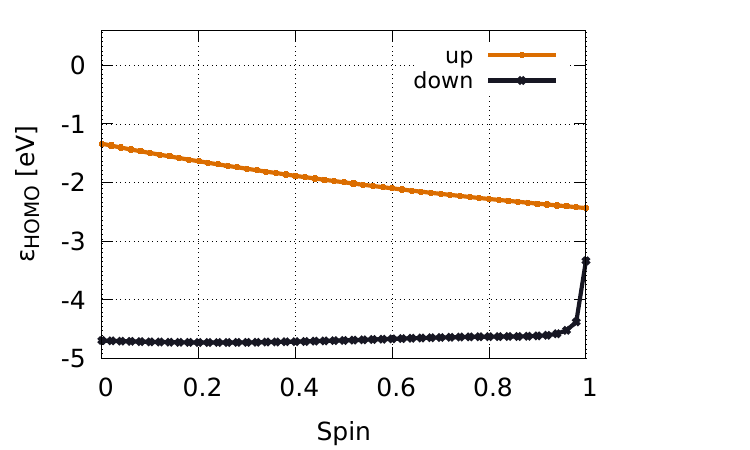}
    \hspace{-12mm}
   \includegraphics[width=0.30\linewidth]{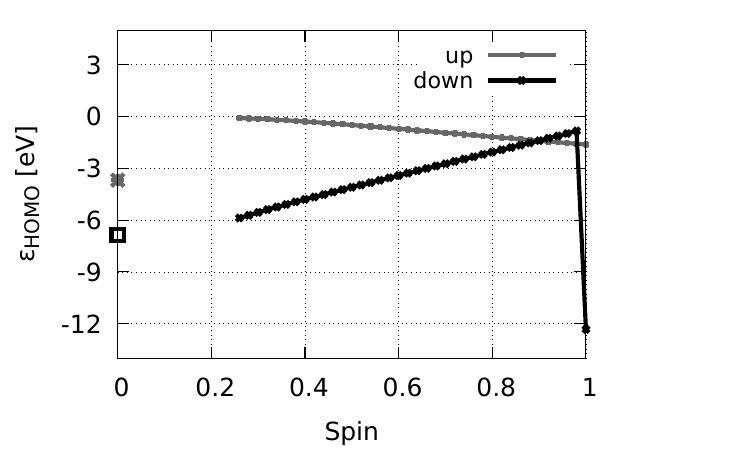}
   \hspace{-12mm}
   \includegraphics[width=0.30\linewidth]{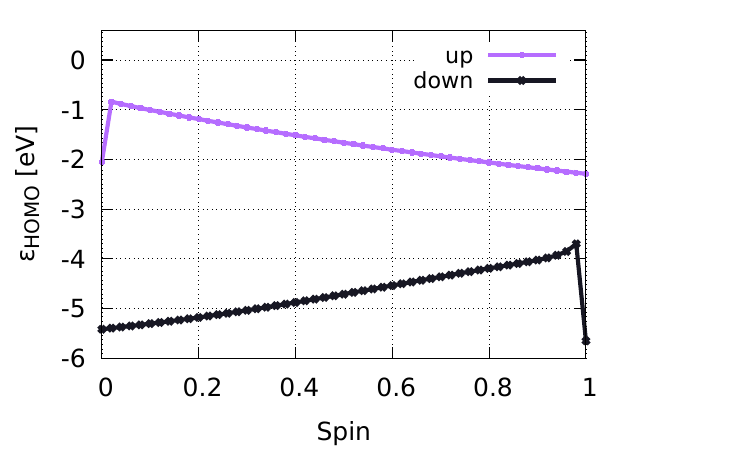}
    \hspace{-12mm}
    \caption{Frontier orbital energies for the Mg atom versus the fractional spin using different XC functionals. From left to right: LSDA, PBE, EXX, PBE0.}
    \label{sfig:Mg_HOMO}
\end{figure*}

\begin{figure*}
    \centering
    \hspace{-12mm}
    \includegraphics[width=0.30\linewidth]{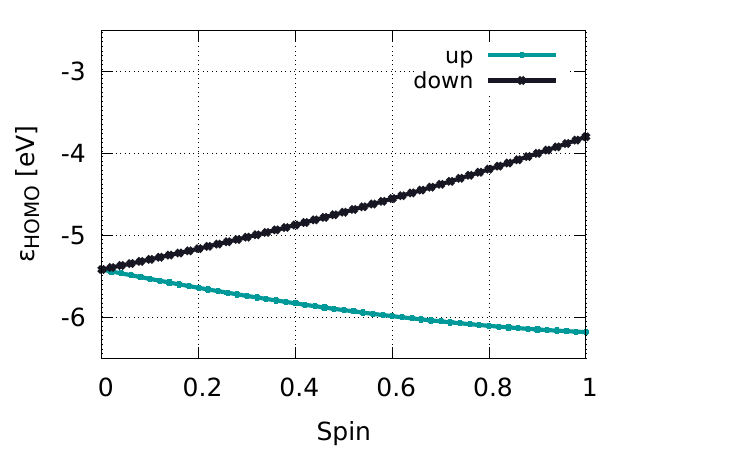}
    \hspace{-12mm}
   \includegraphics[width=0.30\linewidth]{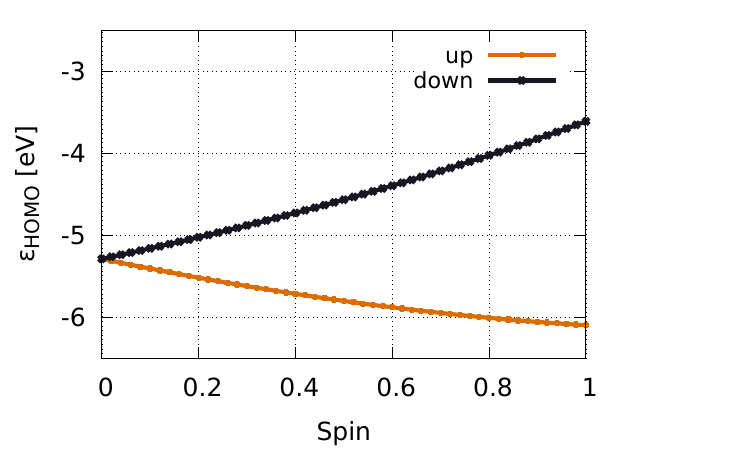}
    \hspace{-12mm}
   \includegraphics[width=0.30\linewidth]{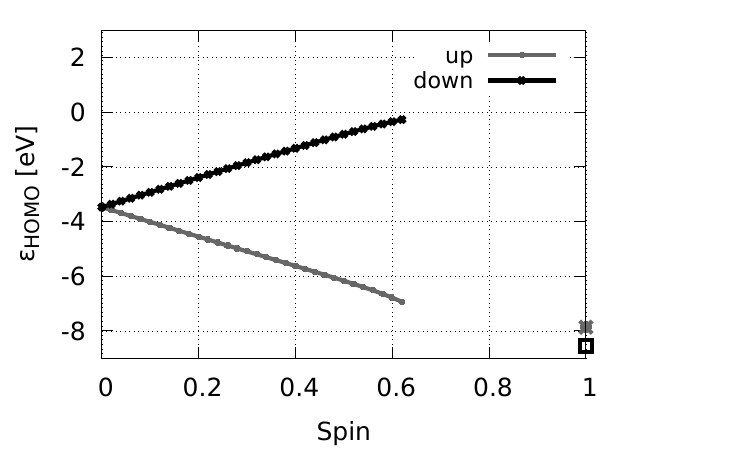}
   \hspace{-12mm}
   \includegraphics[width=0.30\linewidth]{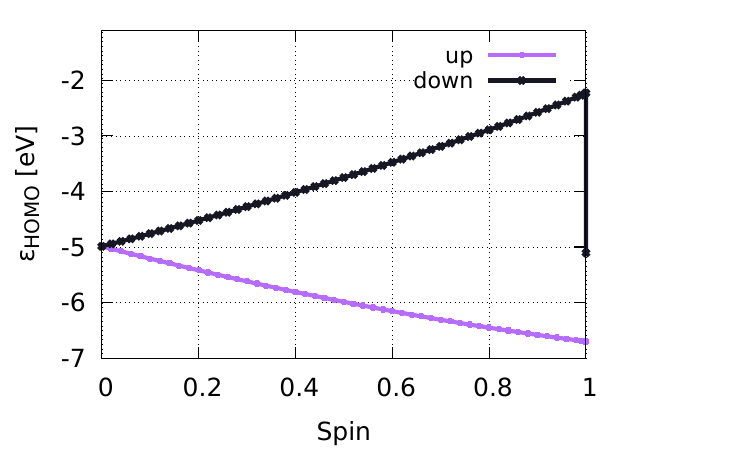}
    \hspace{-12mm}
    \caption{Frontier orbital energies for the C atom versus the fractional spin using different XC functionals. From left to right: LSDA, PBE, EXX, PBE0.}
    \label{sfig:C_HOMO}
\end{figure*}

\begin{figure*}
    \centering
    \hspace{-12mm}
    \includegraphics[width=0.30\linewidth]{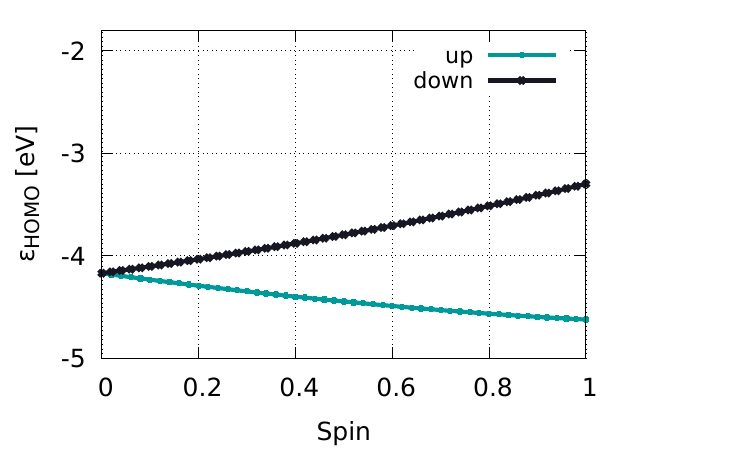}
    \hspace{-12mm}
   \includegraphics[width=0.30\linewidth]{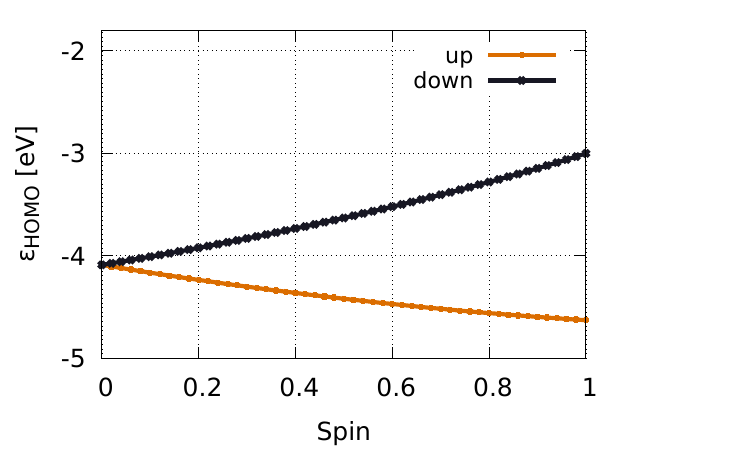}
    \hspace{-12mm}
   \includegraphics[width=0.30\linewidth]{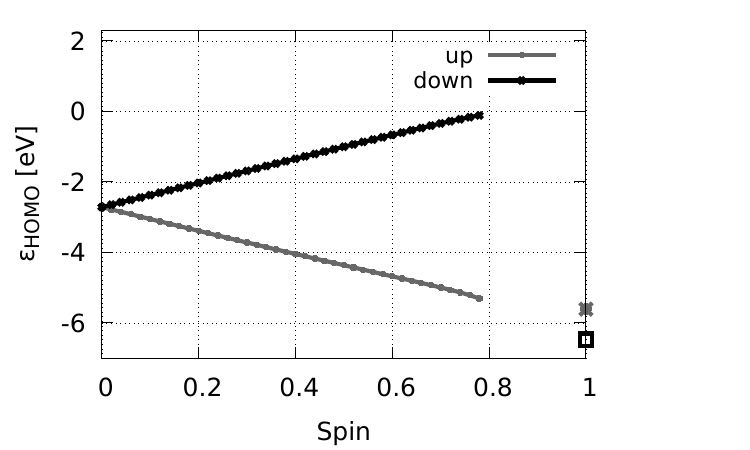}
   \hspace{-12mm}
   \includegraphics[width=0.30\linewidth]{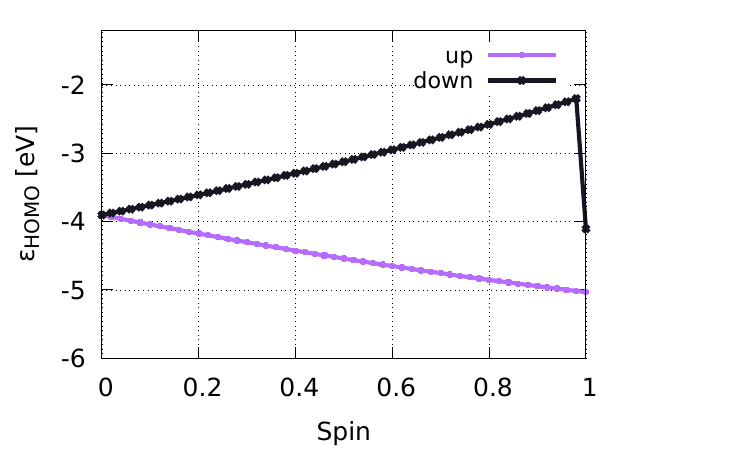}
    \hspace{-12mm}
    \caption{Frontier orbital energies for the Si atom versus the fractional spin using different XC functionals. From left to right: LSDA, PBE, EXX, PBE0.}
    \label{sfig:Si_HOMO}
\end{figure*}

Figure \ref{sfig:Li_HOMO_LDA0} presents the frontier orbital energies for the Li atom using LSDA0 functional. Additional examples of energy deviation, $\Delta E$, versus the fractional spin, $S$, for Case I are the B, K, Be, Mg, C and Si atoms, shown in Fig.~\ref{sfig:Be_Mg_C_SiTotal_E}. The corresponding frontier orbitals are shown in Figs.~\ref{sfig:Be_HOMO} $-$~\ref{sfig:Si_HOMO}.

\newpage

\subsection{CASE II} \label{SI:sec:CaseII}
Figure~2 of the main text presents $\Delta E(S)$ for He, Li$^+$ and Be$^{2+}$. Also in this case,  taking the linear
combination of the EXX and PBE deviations, with weights of $\tfrac{1}{4}$ and $\tfrac{3}{4}$, respectively, provides a close approximation to the PBE0 value, which lies within 0.02 eV, for these three systems, for all values of $S$. 

\begin{figure*}[h]
\centering
\hspace{-12mm}
\includegraphics[width=0.30\linewidth]{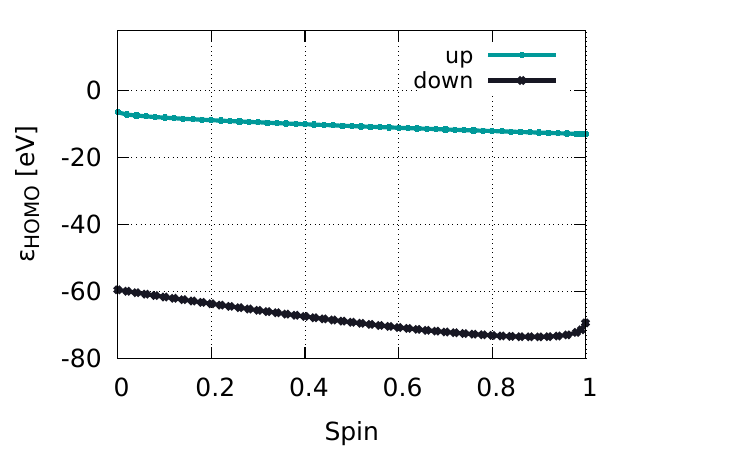}
\hspace{-12mm}
\includegraphics[width=0.30\linewidth]{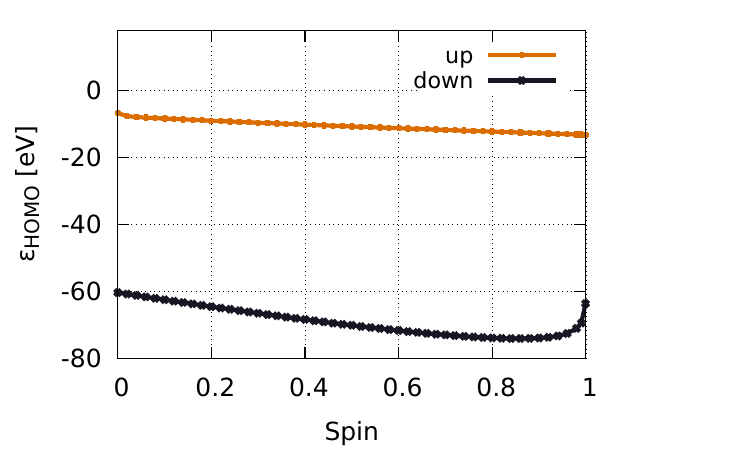}
\hspace{-12mm}
\includegraphics[width=0.30\linewidth]{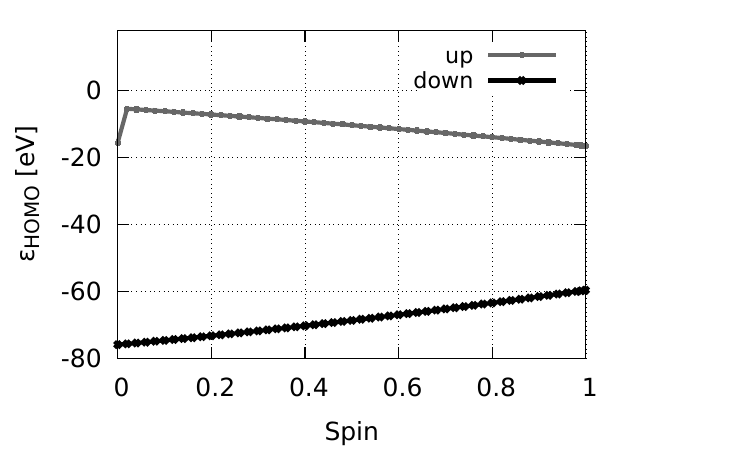}
\hspace{-12mm}
\includegraphics[width=0.30\linewidth]{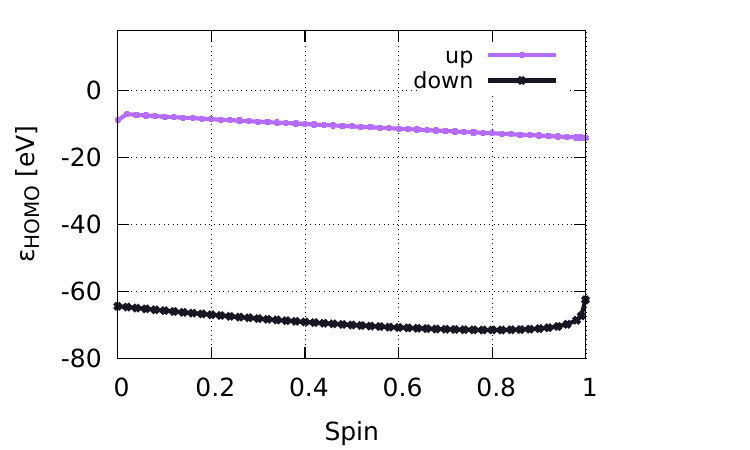}
\hspace{-12mm}
\caption{Frontier orbital energies for Li$^+$, versus the fractional spin, using different XC functionals. From left to right: LSDA, PBE, EXX, PBE0.}
\label{sfig:Li+_HOMO}
\end{figure*}

\begin{figure*}[h]
    \centering
    \hspace{-12mm}
    \includegraphics[width=0.30\linewidth]{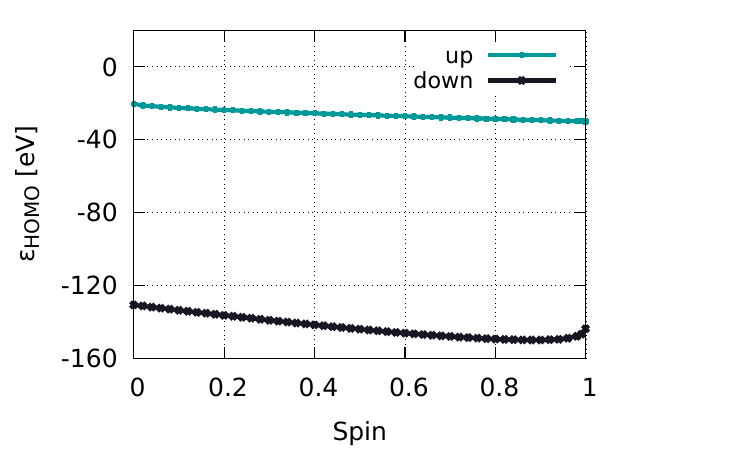}
    \hspace{-12mm}
   \includegraphics[width=0.30\linewidth]{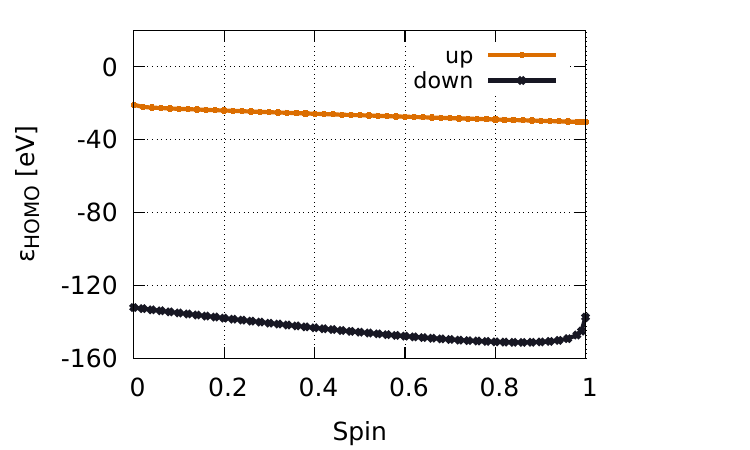}
    \hspace{-12mm}
   \includegraphics[width=0.30\linewidth]{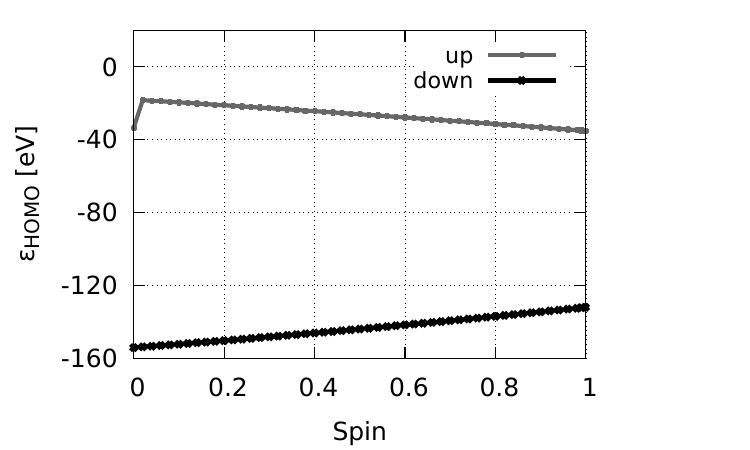}
   \hspace{-12mm}
   \includegraphics[width=0.30\linewidth]{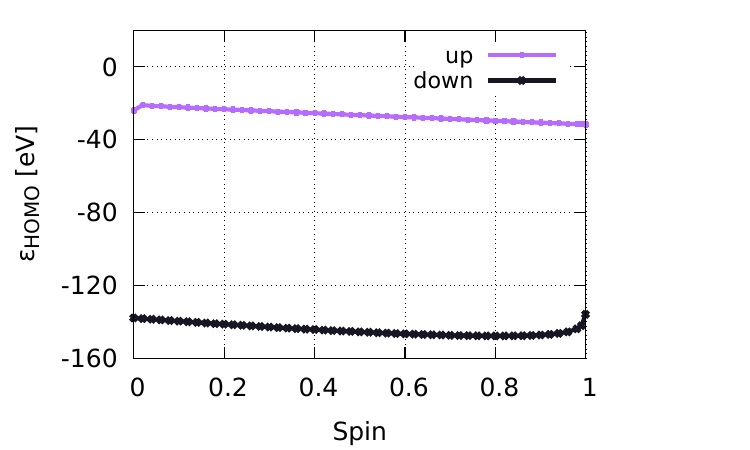}
    \hspace{-12mm}
    \caption{Frontier orbital energies for Be$^{2+}$, versus the fractional spin, using different XC functionals. From left to right: LSDA, PBE, EXX, PBE0.}
    \label{sfig:Be++_HOMO}
\end{figure*}

\begin{figure*}[h]
    \centering
    \hspace{-10mm}
    \includegraphics[width=0.49\linewidth]{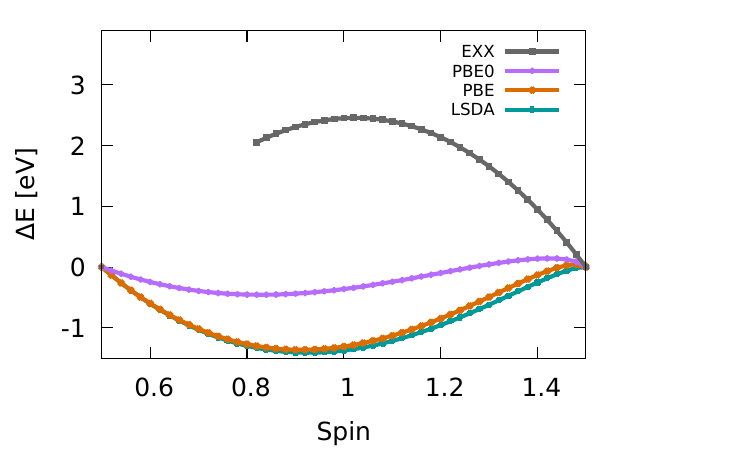}
    \hspace{-10mm}
    \caption{Deviation in energy, $\Delta E$, versus the fractional spin, $S$, for the Li atom, for different XC approximations (see Legend). The spin varies between $\tfrac{1}{2}$ and $\tfrac{3}{2}$.}
    \label{sfig:Li_ex_etot}
\end{figure*}

\begin{figure*}[h!]
    \centering
    \hspace{-12mm}
    \includegraphics[width=0.30\linewidth]{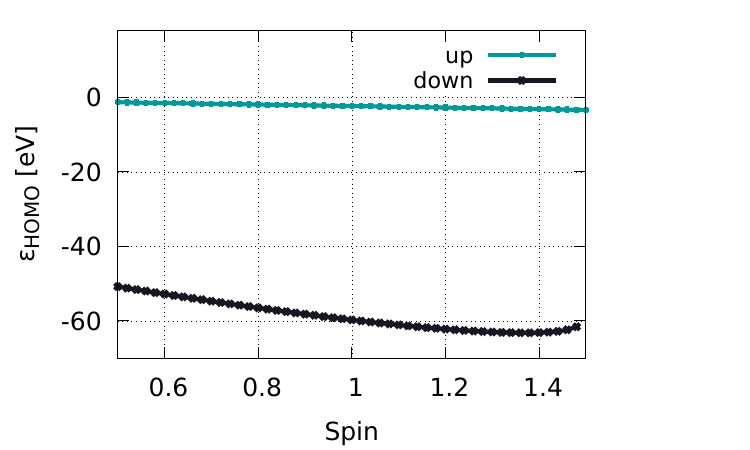}
    \hspace{-12mm}
   \includegraphics[width=0.30\linewidth]{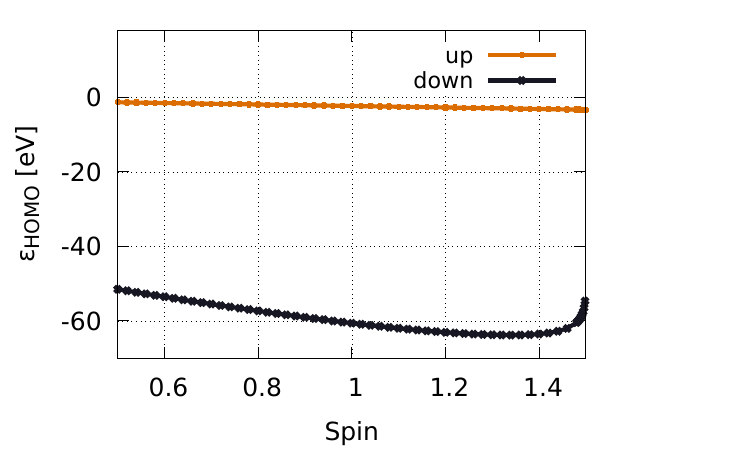}
    \hspace{-12mm}
   \includegraphics[width=0.30\linewidth]{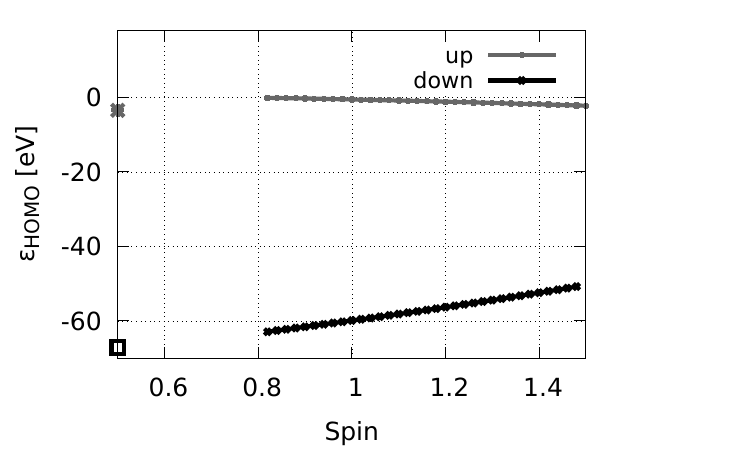}
   \hspace{-12mm}
   \includegraphics[width=0.30\linewidth]{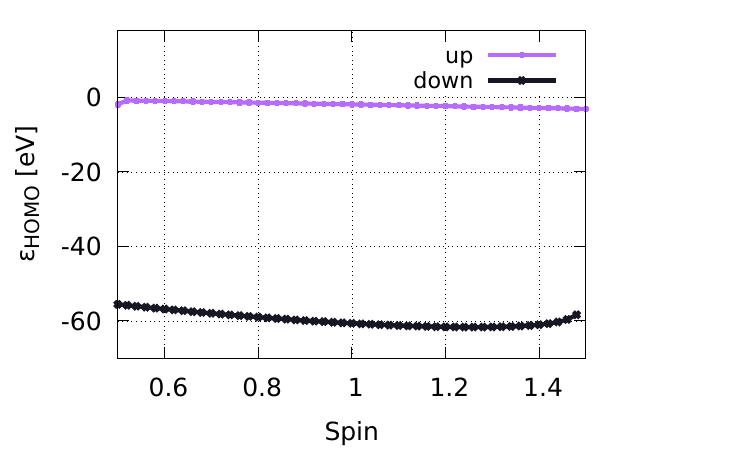}
    \hspace{-12mm}
    \caption{Frontier orbital energies for the Li atom versus the fractional spin, using different XC functionals. From left to right: LSDA, PBE, EXX, PBE0. The spin varies between $\tfrac{1}{2}$ and $\tfrac{3}{2}$.}
    \label{sfig:Li_ex_HOMO}
\end{figure*}

The frontier orbital energies of Li$^+$ and Be$^{2+}$ are shown in Figures \ref{sfig:Li+_HOMO} and \ref{sfig:Be++_HOMO}, respectively. An additional example for Case II behavior is presented for the Li atom, where $S$ changes from $S=0.5$ to $S=1.5$. $\Delta E(S)$ vs.\ the spin and the corresponding frontier orbitals are presented in Figs.~\ref{sfig:Li_ex_etot} and~\ref{sfig:Li_ex_HOMO}, respectively. 

\vspace{-5mm}
\vspace{-5mm}
\vspace{-5mm}
\newpage

\subsection{CASE III} \label{SI:sec:CaseIII}
\begin{figure*}[h]
    \centering
    \hspace{-12mm}
    \includegraphics[width=0.30\linewidth]{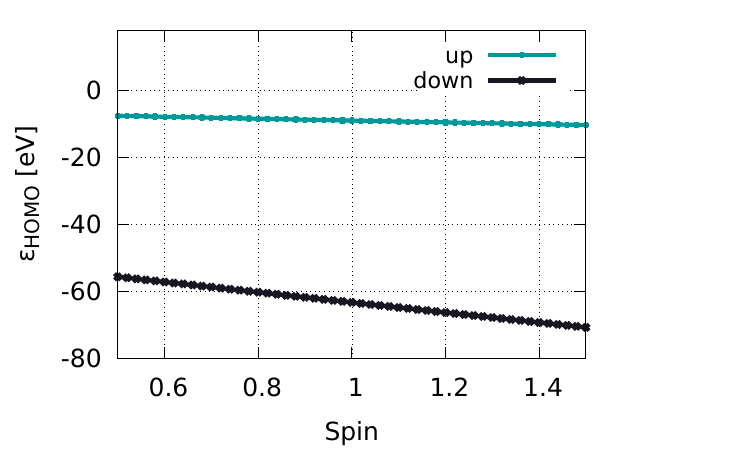}
    \hspace{-12mm}
   \includegraphics[width=0.30\linewidth]{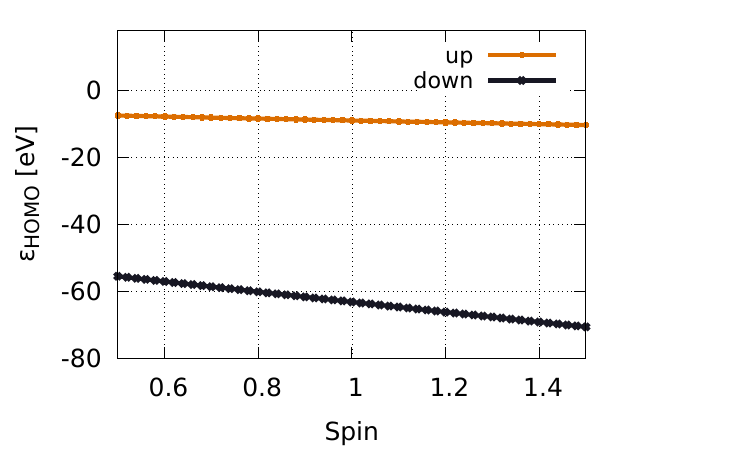}
    \hspace{-12mm}
   \includegraphics[width=0.30\linewidth]{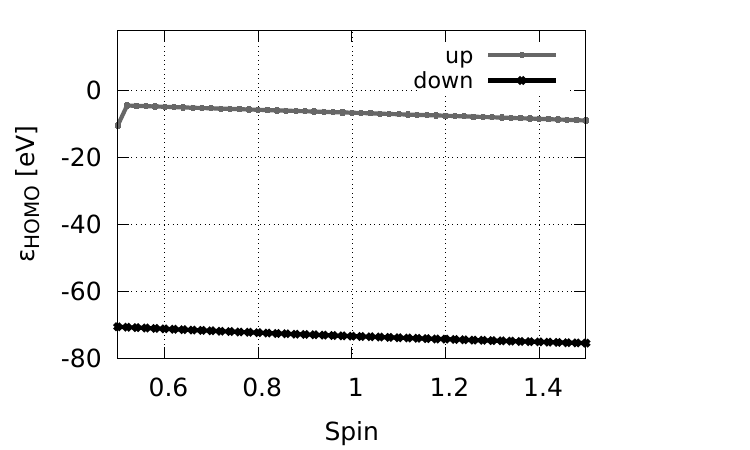}
   \hspace{-12mm}
   \includegraphics[width=0.30\linewidth]{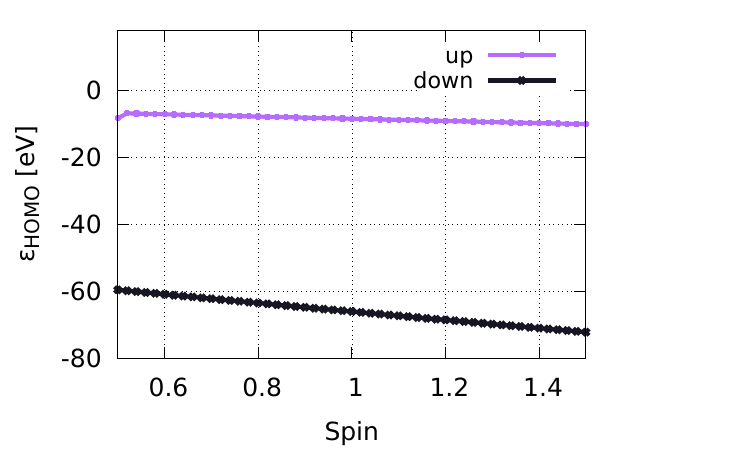}
    \hspace{-12mm}
    \caption{Frontier orbital energies for the Mg$^+$ atom versus the fractional spin, using different XC functionals. From left to right: LSDA, PBE, EXX, PBE0. The spin varies between $\tfrac{1}{2}$ and $\tfrac{3}{2}$.}
    \label{sfig:Mg+_ex_HOMO}
\end{figure*}

Frontier orbital energies of Mg$^+$ are shown in Figure~\ref{sfig:Mg+_ex_HOMO}. 

